\documentclass[letterpaper, 10 pt, conference]{ieeeconf}  

\IEEEoverridecommandlockouts                              
\newcommand{\Tsup}{T^{\mathrm{sup}}}

\newcommand{\mdot}{\dot{m}}
\newcommand{\NTU}{\mathrm{NTU}}

\newcommand{\UA}{U_i A^{\mathrm{rad}}_i}
\newcommand{\cp}{c_{\mathrm{p}}^\mathrm{wat}}

\usepackage{cite}

\title{\LARGE \bf
Model Predictive Control of Heating Systems \\ via Koopman-Operator Bilinearization
}

\author{Paula Chanfreut, Erfan Shakhesi, Mircea Lazar, and Maurice Heemels  \thanks{This research is supported by the Netherlands Enterprise Agency [project number MOOI224004] and by the Dutch Research Council (NWO) under the VENI TTW Grant 21957 ``Dynamic Home Energy Management: Optimizing for a Sustainable Future (DHEMOS)'' [grant ID \protect\url{https://doi.org/10.61686/MVDUJ87852}].}
\thanks{Paula Chanfreut, Erfan Shakhesi, and Maurice Heemels are with the Control Systems Technology Section, Department of Mechanical Engineering,  Eindhoven University of Technology, 5600 MB Eindhoven, The Netherlands (e-mail: p.chanfreut.palacio@tue.nl; e.shakhesi@tue.nl; m.heemels@tue.nl).}
\thanks{Mircea Lazar is with the Control Systems Section, Department of Electrical Engineering, Eindhoven University of Technology, 5600 MB Eindhoven, The Netherlands (e-mail: m.lazar@tue.nl).}
}

\usepackage{graphicx}
\usepackage{amsmath}

\usepackage{amssymb}
\usepackage{siunitx}
\usepackage{float}

\usepackage{booktabs}
\usepackage{tabularx}
\usepackage{siunitx}
\usepackage{mathtools}
\usepackage{bbm}
\usepackage[bb=boondox]{mathalfa}
\usepackage{subcaption}

\DeclareSIUnit\bar{bar}
\usepackage{url}
\allowdisplaybreaks
\usepackage[table]{xcolor}
\newcommand{\rowline}{\arrayrulecolor{gray!50}\specialrule{0.15pt}{0pt}{0pt}\arrayrulecolor{black}}

\usepackage{array}
\newcolumntype{M}[1]{>{\raggedright\arraybackslash}m{#1}}

\usepackage{amsthm} 
\newtheoremstyle{ieeebold}
  {\topsep}{\topsep}
  {\itshape}     
  {}
  {\bfseries}    
  {:}            
  { }
  {\thmname{#1}\thmnumber{ #2}\thmnote{ (#3)}} 

\newenvironment{proofof}[1]
  {\par\noindent\textit{Proof of #1:}\ }
  {\hfill$\blacksquare$\par\medskip}

  \theoremstyle{ieeebold}

\newtheorem{lemma}{Lemma}
\newtheorem{proposition}{Proposition}
\newtheorem{remark}{Remark}
\newtheorem{corollary}{Corollary}
\newtheorem{assumption}{Assumption}

\begin{document}

\maketitle
\thispagestyle{empty}
\pagestyle{empty}

\begin{abstract} 
Hydronic heat emitters are widely used in buildings, but their control is challenging due to nonlinear dynamics, arising from both the heat transfer in each emitter and the hydraulic coupling among those that share a circulating pump. These nonlinearities are often simplified or neglected in the literature. In this paper, we use Koopman theory to make the resulting model more tractable for control. In particular, we propose an input reparameterization and construct an infinite-dimensional set of observables that yields an exact bilinear representation of the system in the chosen lifted space. We then show that truncating the infinite-dimensional representation leads to bounded approximation errors that decrease as the truncation order increases, as well as bounded trajectory deviations over a finite horizon. The resulting bilinear model is incorporated into a model predictive control (MPC) framework to provide thermal comfort in multi-zone buildings while reducing energy use. Using a numerical case study of a five-story building with~20 zones, we show that the proposed Koopman-based MPC achieves a realized performance cost close to that of nonlinear MPC with faster solve times, and outperforms MPC formulations based on time-invariant and time-varying linearizations. 
\end{abstract}

\section{INTRODUCTION} \label{sec:introduction}

Buildings are increasingly required to provide thermal comfort while reducing their energy consumption and environmental footprint~\cite{eea_energy_buildings}. In many buildings, space heating is provided by hydronic heat emitters, such as radiators~\cite{martinopoulos2018comparative}. These heat emitters are supplied with hot water from a heat source, such as a boiler or a heat pump, by means of a circulating pump, and are equipped with valves to regulate their water flow rates. In multi-zone buildings, a single pump typically serves emitters located in different zones. Since the zones generally have different heating demands, the available water flow must be distributed among them accordingly, which gives rise to a flow allocation problem~\cite{li2021study, liu2022thermal}.  A key challenge is that emitters connected to the same pump are hydraulically coupled, which means that the water flow rate through each emitter cannot be adjusted independently by its valve. For example, a fully open valve in one zone may reduce the flow available to other zones, which can cause comfort violations~\cite{shakhesi2026modelpredictivecontroldynamic, en12173215}. Moreover, heat emitters located farther from the pump are naturally affected by a higher hydraulic resistance and may receive less flow than those closer to it~\cite{hakansson2025model}. Consequently, providing thermal comfort in multi-zone buildings requires coordinating the valves in response to the real-time heating demand while explicitly accounting for the hydraulic coupling.

A potential solution is to use online optimization-based control. Specifically, model predictive control (MPC) is a promising approach, as it optimizes the predicted system behavior according to a predefined objective while satisfying the system constraints~\cite{camacho2026mpc}. Although MPC has been widely used in building heating systems~\cite{survey_MPC2025, 7087366, Castilla2014, EMPC_radiator, 5299104, 6573317, 5739562}, its application to this flow allocation problem, also referred to as dynamic hydraulic balancing, remains limited. One reason is that the heat output of hydronic emitters is inherently nonlinear. Using the effectiveness-NTU method~\cite{bergman2011fundamentals}, which is commonly applied to model radiators and radiant panels~\cite{hakansson2025model, wang2025physics}, the heat output is the product of its flow rate, the temperature difference between the supply water and the room air, and an effectiveness that itself depends on the flow rate. 
Accounting for hydraulic coupling introduces additional nonlinearities, since the heat output of each emitter depends on the valve positions in all zones. The resulting MPC problem is nonconvex and grows with the number of zones and the prediction horizon, so it can become computationally demanding in large buildings, and solvers may return only locally optimal solutions. 
A second challenge is that the parameters of the prediction model, including the thermal resistances and  heat transfer coefficients, are often not accurately known in practice. Design data from construction plans frequently deviate from the actual installation, so these parameters must be identified from operational data, and modeling errors can directly degrade the controller performance~\cite{hakansson2025model}.

In this context, Koopman operator theory offers a framework to recast nonlinear dynamics into (bi)linear representations in a lifted space that are more tractable for MPC~\cite{koopman}. This is achieved by describing the nonlinear system through \emph{observables}, i.e.,  functions of the state, whose time evolution is governed by the Koopman operator. For continuous-time control-affine systems, this bilinear representation is exact if the span of the chosen observables is invariant under the Koopman generators associated with the drift and the control inputs~\cite{mauroy2020koopman}. However, such observables are difficult to find in general, especially when the dynamics are non-affine in the input~\cite{IACOB2024111525, 11720155, shang2026existencekoopman}, as is the case of hydronic heating systems.  A common approach in the literature is extended dynamic mode decomposition (EDMD), which approximates the Koopman operator on the span of a predefined dictionary of candidate observables~\cite{williams2015data, koopman}. However, the choice of the dictionary, e.g., polynomial, radial basis, or Fourier functions, is itself nontrivial. Moreover, simply increasing the dictionary size does not necessarily yield a better approximation, while it increases the dimension of the lifted state and the number of coefficients to be learned from data, which is particularly problematic when the available data are limited and the model parameters are uncertain~\cite{haseli2023generalizing}. Instead, this article exploits the structure of the governing equations to derive observables analytically. Such physics-based embeddings avoid the selection of a generic dictionary, retaining a physical interpretation and reducing the identification problem to the estimation of physically meaningful~parameters.

We adopt a model of buildings with hydronic heat emitters that captures both the thermal dynamics and the hydraulic behavior of the valves and circulating pump~\cite{shakhesi2026modelpredictivecontroldynamic}. The model is non-affine in the control inputs and cannot be directly lifted to a bilinear representation using existing Koopman results for control-affine systems. The main contributions of this paper are as follows:
\begin{enumerate}
    \item We introduce an input reparameterization to obtain a control-affine formulation and analytically construct an infinite-dimensional set of observables whose span is invariant under the associated Koopman generators, yielding an exact bilinear representation 
    in the lifted space. Trajectory equivalence between the original nonlinear system and its lifted representation is established in both the forward and reverse directions.
    
    \item For practical implementation, we construct a finite-dimensional truncation of the lifted system and derive explicit bounds on the truncation error, showing that it decreases as the truncation order increases. A bound on the deviation between trajectories of the truncated lifted system and those of the original nonlinear system over a finite horizon is also derived.
    \item Finally, we integrate the truncated lifted model into  MPC to jointly optimize the supply temperature and valve positions, performing dynamic hydraulic balancing to provide multi-zone thermal comfort while reducing energy use. The proposed Koopman-based MPC achieves a lower closed-loop cost than linear time-invariant and linear time-varying MPC with comparable computation times, and performance close to that of nonlinear MPC at a lower computational cost.
\end{enumerate}

The remainder of this paper is organized as follows. Section~\ref{sec:modeling} introduces the building model. Section~\ref{sec:koop} presents the proposed analytical Koopman bilinear embedding and the associated truncated model. Section~\ref{sec:controller_design} describes its integration into MPC, and Section~\ref{sec:NCS} presents the associated numerical results. Finally, Section~\ref{sec:conclusion} concludes the paper and outlines directions for future work.
\section{Modeling} \label{sec:modeling}
\begin{table}[!b]
\centering
\caption{Thermal model parameters and variables.}
\label{tab:symbols_thermal_model}
\setlength{\tabcolsep}{6pt}
\renewcommand{\arraystretch}{1.15}
\begin{tabularx}{\linewidth}{@{} l l X @{}}\toprule
\textbf{Symbol} & \textbf{Unit} & \textbf{Description} \\ 
\midrule
$C_{\mathrm{z},i}$ & $\si{\joule\per\kelvin}$ & Heat capacity of zone $i$ \\ \rowline

$C_{\mathrm{w},i}$ & $\si{\joule\per\kelvin}$ & Heat capacity of wall $i$ \\ \rowline

$T_{\mathrm{z},i}$ & $\si{\kelvin}$ & Temperature of zone $i$ \\ \rowline

$T_{\mathrm{w}, i}$ & $\si{\kelvin}$ & Temperature of wall $i$ \\ \rowline


$T_{\mathrm{o}}$ & $\si{\kelvin}$ & Outdoor temperature \\ \rowline



$A_{\mathrm{w},i}$ & $\si{\square\meter}$ & Area of wall $i$, excluding window areas \\ \rowline

$A_{\mathrm{wd},i}$ & $\si{\square\meter}$ & Area of window(s) located on wall $i$ (if none, $A_{\mathrm{wd},i} = 0$) \\ \rowline

$R_{\mathrm{w},i}$ & $\si{\kelvin \square \meter \per\watt}$ & Thermal resistance of wall $i$ \\ \rowline

$R_{\mathrm{wd},i}$ & $\si{\kelvin \square \meter\per\watt}$ & Thermal resistance of window(s) on wall $i$ \\ \rowline


$\dot{m}_{i}$ & $\si{\kilogram \per \second}$ & Water mass flow rate through emitter~$i$ \\ \rowline

$\mathcal{N}_{\mathrm{z},i}$ & -- & Set of indices of zones adjacent to wall $i$; index~$0$ corresponds to the outdoor environment \\ \rowline

$\mathcal{N}_{\mathrm{w},i}$ & -- & Set of indices of walls surrounding zone $i$ \\ \rowline


$G^{\mathrm{inf}}_i$ & $\si{\watt\per\kelvin}$ & Infiltration conductance of zone $i$ \\ \rowline

$Q^{\mathrm{rad}}_i$ & $\si{\watt}$ & Heat transfer rate supplied by emitter $i$ \\ \rowline

$Q^{\mathrm{int}}_i$ & $\si{\watt}$ & Internal heat gain rate in zone $i$ \\ \rowline


$Q^\mathrm{sol}_{\mathrm{z},i}$ & $\si{\watt}$ & Solar heat gain in zone $i$ \\ \rowline

$Q^\mathrm{sol}_{\mathrm{w},j}$ & $\si{\watt}$ & Solar heat gain on wall $j$ \\ \rowline




$n_{\mathrm{z}}$ & -- & Number of zones \\ \rowline

$n_{\mathrm{w}}$ & -- & Number of walls \\ 
\bottomrule
\end{tabularx}
\end{table}
This section presents the modeling of a multi-zone building with heat emitters.
\subsection{Thermal model}
The dynamics of air temperature of zone $i$, i.e., $T_{\text{z},i}$, ~\mbox{$i \in \{1,2,\hdots, n_{\mathrm{z}}\}$}, is described by~\cite{BRCM_toolbox}:
\begin{align}\label{eq:zone_base}
    &C_{\text{z},i} \frac{dT_{\text{z},i}}{dt} =  \sum_{j \in \mathcal{N}_{\mathrm{w},i}}\frac{A_{\text{w},j}}{R_{\text{w},j}}\left(T_{\text{w}, j} - T_{\mathrm{z},i}\right) + Q^\text{rad}_i + Q^\text{sol}_{\text{z},i}  \nonumber \\
    &+\sum_{j \in\mathcal{N}_{\mathrm{w},i}}\frac{A_{\text{wd},j}}{R_{\text{wd},j}}(T_{\text{o}} - T_{\text{z},i}) + G^\text{inf}_i (T_\mathrm{o} - T_{\mathrm{z},i}) + Q^\text{int}_i,  
\end{align}
where all variables are defined in Table~\ref{tab:symbols_thermal_model}. For ease of notation, we omit the time argument $t$ in \eqref{eq:zone_base} and also throughout this section. The terms on the right-hand side of~\eqref{eq:zone_base} represent, respectively, heat transfer between the zone air and adjacent walls, heat supplied by the heat emitters, solar heat gains, heat transfer to the outdoor environment through the windows, heat transfer due to outdoor air infiltration, and internal heat gains (i.e., heat released inside the zone by occupants, lighting, appliances, etc.). The outdoor temperature $T_{\text{o}}$ (prediction or measurement) is obtained from a weather forecasting station,  $Q^{\text{int}}_{i}$ is estimated (e.g., from historical data or room booking information), and $Q^\text{sol}_{\mathrm{z},i}$ is computed as described in~\cite{shakhesi2026modelpredictivecontroldynamic}. These three variables are treated as known exogenous inputs, while $Q^{\text{rad}}_i$ is the only (indirectly) controllable~variable. Note that the superscript ``rad'' is used since radiators are the heat emitters considered in the case study. Nonetheless,  the proposed approach applies equally to any heat emitter that can be modeled as described in Section~\ref{sec:radiator_heating}. 

\subsubsection{Wall temperatures}
To characterize model~\eqref{eq:zone_base}, we require temperatures $T_{\mathrm{w}, j}$ for $j\in \mathcal{N}_{\mathrm{w}, i}$ and $i\in \{1, 2, ..., n_\mathrm{z}\}$, which represent the mid-plane temperatures of each wall~$j$. Following \cite{BRCM_toolbox}, these dynamics can be written as
\begin{align}
    &C_{\mathrm{w},j} \frac{dT_{\mathrm{w},j}}{dt} =  \sum_{i \in \mathcal{N}_{\mathrm{z}, j}} \frac{A_{\mathrm{w}, j}}{R_{\mathrm{w}, j}}( T_{\mathrm{z},i} - T_{\mathrm{w}, j}) + Q^\text{sol}_{\mathrm{w}, j},\label{eq:wall_temp}
\end{align}
where $T_{\mathrm{z},0} \coloneqq T_{\mathrm{o}}$. On the right-hand side of \eqref{eq:wall_temp}, the first term models the heat transfer between the wall and its adjacent zones, and the second term accounts for solar radiation absorbed by wall $j$~\cite{shakhesi2026modelpredictivecontroldynamic}. 

Throughout, we consider the following physically motivated assumption and remark:
\begin{assumption}\label{ass:wall_slow}
The wall temperatures evolve on a significantly slower time scale than zone air temperatures. For the MPC design, they can be treated as exogenous signals $\hat T_{\mathrm{w},j}$,  $j \in \{1, 2, \dots,n_\mathrm{w}\}$, over the prediction horizon, rather than jointly-optimized~states. 
\end{assumption}
\begin{remark}
In \eqref{eq:zone_base} and \eqref{eq:wall_temp}, the term ``walls'' also includes floor and ceiling elements.
\end{remark}

\subsubsection{Heat emitters}\label{sec:radiator_heating}
Heat output $Q^\mathrm{rad}_i$, $i\in \{1, 2, ..., n_\mathrm{z}\}$, is modeled using the effectiveness--NTU method~\cite{bergman2011fundamentals} as
\begin{equation}
    Q^\text{rad}_i = \mdot_i c_{\mathrm{p}}^\mathrm{wat} \bigl(\Tsup-T_{\mathrm{z}, i}\bigr)
        \bigl(1-e^{-\NTU_i}\bigr),
    \label{eq:rad_dynamics}
\end{equation}
where $$\NTU_i = \dfrac{c_i}{\dot m_i}, \text{   with   } c_i = \dfrac{\UA}{c_{\mathrm{p}}^{\mathrm{wat}}}.$$ 
Here, $\UA$ is the overall heat emitter conductance~[W/K], $c_{\mathrm{p}}^{\mathrm{wat}}$ is the specific heat capacity of water $[\si{\joule \per \kilogram \per \kelvin}]$,~$\mdot_i$ denotes the mass flow rate [kg s$^{-1}$],  and $T^{\text{sup}}$  is a control input representing the water supply temperature $[\si{\kelvin}]$. The latter two variables are physically constrained as
\begin{equation}\label{eq:const_mT}
\begin{split}
    &\dot m_{\min,i} \leq   \dot m_i \leq  \dot m_{\max,i}, \\
    &T^{\text{sup}}_{\min} \leq   T^{\text{sup}} \leq T^{\text{sup}}_{\max}, 
\end{split}
\end{equation}
where $\dot m_{\min,i}, \dot m_{\max,i}, T^{\text{sup}}_{\min}, T^{\text{sup}}_{\max} \in \mathbb{R}_{>0}$, $i\in \{1, 2, ..., n_\mathrm{z}\}$. For notational convenience, let us use $u_\mathrm{T} = T^{\text{sup}}$ in what follows.

\begin{assumption}\label{ass:parallel_radiators}
The emitters are installed in parallel (see Figure~\ref{fig:NCS_building}), and heat losses in the connecting pipes are negligible. Consequently, the supply water temperature is  the same for all of them, making it a shared control decision.
\end{assumption}

\subsection{Hydraulic model}\label{sec:hydraulic_model} The computation and optimization of flow rates~$\dot{m}_{i}$ for all $i\in \{1, 2,..., n_\mathrm{z}\}$ require a model of the heating system hydraulic coupling. We consider a parallel configuration including a set of heat emitters, a heat source, and a circulating pump. For more general models of interconnected hydraulic networks, refer to \cite{5682372, 10443253}. The pump head $H \; [\si{\meter}]$ can be approximated using manufacturer datasheet information (typically provided as the $H-Q$ curve) as a quadratic function of the overall mass flow rate $ M = \sum_{i=1}^{n_{\mathrm{z}}} \dot{m}_{i}$: 
\begin{align}
    H = a_{\mathrm{p}} - b_{\mathrm{p}}M^2, \label{eq:pump-head}
\end{align}
where $a_{\mathrm{p}}, b_{\mathrm{p}} \in \mathbb{R}$ are curve-fitting coefficients. Each emitter experiences approximately the same pressure drop $\Delta p \; [\si{\pascal}]$, which can be written as
\begin{align}\label{eq:deltap}
    \Delta p = \kappa_i \dot{m}_{i}^2 = \kappa_j \dot{m}_{j}^2, \ \ \ i, j \in \{1, 2, ..., n_\mathrm{z} \}, 
\end{align}
where $\kappa_i \; [\si{\pascal \square \second \per \square \kilogram}]$ denotes the hydraulic resistance of heat emitter~$i$, computed as~\cite{AGNER2024133462}:
\begin{align}
\kappa_i
&=
\bigg(
\frac{\kappa_{\mathrm{v}, i}}{u_{\mathrm{v}, i}} 
\bigg)^2
+
\kappa_{0, i}^2,
\label{eq:hydraulic_resistance}
\end{align}
where $u_{\mathrm{v}, i} \in [0, 1]$ is a control input representing the valve position ($u_{\mathrm{v}, i}=1$ corresponds to fully open and $u_{\mathrm{v}, i}=0$ to fully closed), and
\begin{equation}
    \kappa_{\mathrm{v, i}} = \frac{3600\sqrt{10^5}}{\rho^{\mathrm{wat}}\,k_{\mathrm{v},i}}, \qquad \kappa_{0, i} = \frac{3600\sqrt{10^5}}{\rho^{\mathrm{wat}}\,k_{0,i}}. \label{eq:flow_coefficients}
\end{equation}
Here, $k_{\mathrm{v},i}$ and $k_{0,i} \;
[\si{\cubic\meter\per\hour\per\sqrt{\bar}}]$ are the manufacturer-specified flow coefficient of the fully open valve and the remaining-component flow coefficient (e.g., lockshield valve), respectively. Using~\eqref{eq:deltap}, the overall flow rate $M$ can be written as
\begin{align}
    M   = \sqrt{\Delta p} \underbrace{\sum_{i=1}^{n_{\mathrm{z}}} \frac{1}{\sqrt{\kappa_i}} }_{\coloneqq \kappa_{\sum}}. \label{eq:sum-q}
\end{align}
Converting \eqref{eq:pump-head} to pressure form and substituting \eqref{eq:sum-q} yields
\begin{align}
    \rho^\mathrm{wat} g a_{\mathrm{p}} - \rho^\mathrm{wat} g b_{\mathrm{p}} ( \sqrt{\Delta p} \kappa_{\sum} ) ^2 = \Delta p,\label{eq:delta-p-2}
\end{align}
where $\rho^{\mathrm{wat}} \; [\si{\kilogram \per \cubic \meter}]$ is the density of water (assumed constant), and $g \; [\si{\meter \per \square \second}]$ is the standard acceleration due to gravity. By rewriting \eqref{eq:delta-p-2} and combining it with \eqref{eq:deltap}, flow rate $\dot m_i$, $i \in \{1,2,\hdots, n_{\mathrm{z}}\}$, is given by   
\begin{align}
    \dot{m}_{i}(u_{\mathrm{v}}) = \sqrt{\frac{\rho^\mathrm{wat} ga_{\mathrm{p}}}{1 + \rho^\mathrm{wat} gb_{\mathrm{p}} \left(\kappa_{\sum}(u_\mathrm{v}) \right)^2}} \frac{1}{\sqrt{\kappa_i(u_{\mathrm{v}, i})}}, \label{eq:flow_rate}
\end{align}
where $u_{\mathrm{v}} = [u_{\mathrm{v}, i}]_{i=1}^{n_\mathrm{z}}$ collects the valve positions across all~$n_\mathrm{z}$ zones. The dependency on input vector $u_{\mathrm{v}}$ is made explicit in~\eqref{eq:flow_rate} to highlight a key coupling effect: $\dot m_i$ depends on all valve decisions $[u_{\mathrm{v},j}]_{j=1}^{n_\mathrm{z}}$ through $\kappa_{\sum}(u_\mathrm{v})$, not merely on~$u_{\mathrm{v},i}$. Consequently, changing $u_{\mathrm{v},i}$ affects not only $Q^\text{rad}_i$ but also $Q^\mathrm{rad}_j$ for all $j \neq i$. This cross-coupling is a defining feature of the hydraulic balancing problem.

\subsection{State-space representation}

\vspace{5pt}

The overall system given by~\eqref{eq:zone_base},~\eqref{eq:wall_temp},~\eqref{eq:rad_dynamics}, and~\eqref{eq:flow_rate} can be written in compact form as 
\begin{equation}\label{eq:ss_T}
\dot{T} = f\big(T, u, d\big), 
\end{equation}
where $f: \mathbb{R}^{n_\mathrm{z}+n_\mathrm{w}} \times \mathbb{R}^{n_\mathrm{z}+1} \times \mathbb{R}^{2n_\mathrm{z} + n_\mathrm{w} + 1} \to \mathbb{R}^{n_\mathrm{z}+n_\mathrm{w}}$ is a nonlinear function, and the state and input vectors are respectively given by
\begin{align*}\label{eq:Tandu}
T \coloneqq \begin{bmatrix}
    [T_{\mathrm{z},i}]_{i=1}^{n_{\mathrm{z}}} \\ [T_{\mathrm{w},i}]_{i=1}^{n_{\mathrm{w}}}
\end{bmatrix} \text{   and   } u := \begin{bmatrix} u_\mathrm{T} \\ [u_{\mathrm{v}, i}]_{i=1}^{n_{\mathrm{z}}} \end{bmatrix}.
\end{align*}
Additionally, the exogenous input is 
\begin{align*}
    d := \begin{bmatrix}
        T_{\text{o}} & (Q^\text{int})^{\top} & (Q^{\text{sol}}_\mathrm{z})^{\top} & (Q^{\text{sol}}_\mathrm{w})^{\top}
    \end{bmatrix}^{\top}, 
\end{align*}
where 
\begin{equation*}
Q^{\mathrm{int}} \coloneqq [Q^{\mathrm{int}}_i]_{i = 1}^{n_{\mathrm{z}}}, \  Q^{\mathrm{sol}}_\mathrm{z} \coloneqq [Q^{\mathrm{sol}}_{\mathrm{z}, i}]_{i = 1}^{n_{\mathrm{z}}}, \ Q^{\mathrm{sol}}_\mathrm{w} \coloneqq [Q^{\mathrm{sol}}_{\mathrm{w}, i}]_{i = 1}^{n_{\mathrm{w}}}.    
\end{equation*}
Note that $f$ can be obtained  substituting~\eqref{eq:flow_rate} into~\eqref{eq:rad_dynamics} to express $Q^{\mathrm{rad}}_i$ directly in terms of $u$, and substituting the result into~\eqref{eq:zone_base} together with~\eqref{eq:wall_temp}.

\begin{remark}
    Walls with identical orientation (e.g., surface azimuth) receive the same solar irradiation, so only unique values need to be stored to avoid redundancy.
\end{remark}

\subsubsection{Prediction model}
Under Assumption~\ref{ass:wall_slow}, the prediction model associated with~\eqref{eq:ss_T}  can be abstracted~as:
\begin{equation}\label{eq:nl_dynamics}
\begin{split}
    &\dot x_i = a_i x_i + b_i g_i(x_i, u) + w_i, \\
     &g_i(x_i, u) = \dot m_i(u_\mathrm{v}) \left(u_\mathrm{T} - x_i\right)\left(1-e^{-c_i/\dot m_i(u_\mathrm{v})}\right), \\ &i\in \{1, 2, ..., n_\mathrm{z}\},
    \end{split}
\end{equation}
where $x_i = T_{\mathrm{z},i}$  and function $g_i: \mathbb{R} \times \mathbb{R}^{n_\mathrm{z}+1}\rightarrow\mathbb{R}$ collects the nonlinear terms inherited from the heat exchange behavior~\eqref{eq:rad_dynamics}. Note also that $\dot m_i(u_{\mathrm{v}})$ is itself a function of the entire valve input vector, as shown in~\eqref{eq:flow_rate}.
In addition, $a_i$ is defined from system parameters, $b_i=c_\mathrm{p}^ \mathrm{wat}/C_{\text{z},i}$, and $w_{i}$ aggregates all the effect of the exogenous inputs $T_\mathrm{o}$, $Q^\text{int}_i$, and~$Q^{\text{sol}}_i$, and that of $\hat{T}_{\mathrm{w}, j}$. Specifically,
\begin{equation*}
    \begin{split}
       & a_i =  - \sum_{j \in \mathcal{N}_{\mathrm{w},i}} \left( \frac{A_{\text{w},j}}{C_{\text{z},i} R_{\text{w},j}} + \frac{A_{\text{wd},j}}{C_{\text{z},i} R_{\text{wd},j}} \right) - \frac{G^{\text{inf}}_i}{C_{\text{z},i}}, \\
        &  w_i = \sum_{j \in \mathcal{N}_{\mathrm{w},i}} \left( \frac{A_{\text{w},j}}{C_{\text{z},i} R_{\text{w},j}} \hat{T}_{\text{w}, j} + \frac{A_{\text{wd},j}}{C_{\text{z},i} R_{\text{wd},j}} T_\mathrm{o} \right)  \\  &\qquad + \frac{G^{\text{inf}}_i}{C_{\text{z},i}}T_\mathrm{o} +  \dfrac{1}{C_{\text{z},i} } \left(Q^\text{int}_i + Q^\mathrm{sol}_{\text{z},i} \right). 
    \end{split}
\end{equation*}
As the heat output \eqref{eq:rad_dynamics} and the water flow rate \eqref{eq:flow_rate} introduce nonlinearities, obtaining an exact linear discretization of~\eqref{eq:nl_dynamics} for MPC is infeasible. We therefore seek a bilinear approximation using the Koopman operator framework~\cite{koopman}, which will be compared with that of MPC controllers based on first-order Jacobian linearizations.

\section{Analytical Koopman-Bilinear Embedding} \label{sec:koop}

For a dynamical system with dynamics $\dot x=h(x,u,d)$ as in~\eqref{eq:ss_T}, with state~$x\in\mathbb R^{n_\mathrm{x}}$, input $u\in\mathbb R^{n_\mathrm{u}}$, disturbances $d\in\mathbb R^{n_\mathrm{d}}$, and a continuously differentiable vector field $h:\mathbb R^{n_\mathrm{x}}\times\mathbb R^{n_\mathrm{u}} \times \mathbb R^{n_\mathrm{d}} \to\mathbb R^{n_\mathrm{x}}$, the Koopman operator framework~\cite{koopman} constructs a higher-dimensional representation by lifting the state through a vector of observable functions $\psi:\mathbb{R}^{n_\mathrm{x}}\to\mathbb{R}^N$, where $N\gg n_\mathrm{x}$. This defines the lifted state as
\begin{equation}
z \coloneqq \psi\big(x\big),
\end{equation}
and the lifted dynamics as 
\begin{equation}\label{eq:Koopman_general}
  \dot\psi(x)=(\mathcal K\psi)(x),
  \qquad
    (\mathcal K\psi)(x)\coloneqq\frac{\partial\psi}{\partial x}(x)\,h(x,u, d),
\end{equation}
where $\mathcal K$ is the Koopman generator. The observables $\psi$ are chosen such that the nonlinear dynamics of $x$ are (approximately) represented as (bi)linear dynamics in the lifted space. As mentioned, we seek a model of the~form
\begin{equation}
    \dot z = A z + B \tilde u +  G d +  E (z \otimes \tilde u)  +  F (z \otimes d) \label{eq:lifted_system_cont}
\end{equation}
to capture the dynamics in \eqref{eq:nl_dynamics}.
Here, $\otimes$ denotes the Kronecker product, $\tilde u \in \mathbb{R}^{n_\mathrm{z} +1}$ is the (possibly reparametrized) control input, and $A$, $B$, $G$, $E$, and $F$ are matrices of appropriate dimensions to be identified.

\subsection{Input reparameterization}

Since the dynamics~\eqref{eq:nl_dynamics} are not affine in input $u$, neither is their Koopman generator, so an exact model bilinear in $u$ and~$x$ cannot be obtained by lifting the zone temperatures alone. We therefore use a change of variable. Specifically, let us introduce auxiliary~variable
\begin{equation}\label{eq:log_obs1}
\mu_i \coloneqq \log \dot m_i,
\end{equation}
 for each zone $i \in \{1,2,\dots, n_\mathrm{z}\}$, and treat as control inputs the log-flow-rate derivative $\tilde u_{\mu,i} \coloneqq \dot \mu_i$, together with the supply-temperature rate $\tilde u_\mathrm{T} \coloneqq \dot T^\mathrm{sup}$, i.e.,
\begin{equation}\label{eq:tilde_u}
\tilde u \coloneqq \begin{bmatrix} \tilde u_\mathrm{T} \\ [\tilde u_{\mu,i}]_{i=1}^{n_\mathrm{z}} \end{bmatrix} \in \mathbb{R}^{n_\mathrm{z}+1}.
\end{equation}
The original coupling in~\eqref{eq:flow_rate} will be considered to relate $[\tilde u_{\mu,i}]_{i=1}^{n_\mathrm{z}}$ back to the physical valve inputs $[u_{\mathrm{v},i}]_{i=1}^{n_\mathrm{z}}$. We return to this map in Section~\ref{sec:controller_design}. Define also the augmented state of zone $i$ as $\xi_i = \begin{bmatrix} x_i &  \mu_i &  u_\mathrm{T}\end{bmatrix}^\top$, where $u_\mathrm{T}$ and $\mu_i$ become states through integrators. The dynamics of $\xi_i$ are
\begin{equation}\label{eq:augmented_system}
\dot \xi_i = \begin{bmatrix}
    f_{0,i}(\xi_i) + w_i \\  \tilde u_{\mu,i} \\ \tilde u_T 
\end{bmatrix},
\end{equation} 
where $f_{0,i}(\xi_i) = a_i x_i + b_i g_i(\xi_i)$ is the drift term.  

For any differentiable vector of  observables $\psi_i(\xi_i)$, applying the chain rule, we have that
\begin{equation*}
\begin{split}
        \dot \psi_i(\xi_i) &=    \dfrac{\partial \psi_i}{\partial x_i} \dot x_i +   \dfrac{\partial \psi_i}{\partial \mu_i} \dot \mu_i +   \dfrac{\partial \psi_i}{\partial u_\mathrm{T}} \dot u_\mathrm{T}. 
\end{split}
\end{equation*}
Using notation $\partial_{\xi_j}\psi_i = \partial\psi_i/\partial\xi_j$, the Koopman generator for each zone $i$ is 
\begin{equation}\label{eq:Koopman_generator}
\mathcal{K}_{i} = f_{0,i} \partial_{x_i} + \tilde u_T\, \partial_{u_T} + \tilde u_{\mu,i}\, \partial_{\mu_i} +  w_i\, \partial_{x_i},
\end{equation}
and, therefore, $\dot \psi_i(\xi_i) = (\mathcal{K}_i \psi_i)(\xi_i)$. The extension to the global system holds simply by aggregating the augmented states across all zones $i\in \{1, 2, ..., n_\mathrm{z}\}$.

\subsection{Dictionary selection}

For every~$i\in \{1, 2, ..., n_\mathrm{z}\}$, let us build observables $\psi_i(\xi_i)$ using the following variables:
\begin{subequations}\label{eq:observables}
    \begin{align}
\phi_{i,p} &:= e^{(1-p)\mu_i}, \label{eq:phi_def}\\
\theta_{i,p} &:= \phi_{i,p}\,x_i, \label{eq:theta_def} \\
\chi_{i,p} &:= \phi_{i,p} \bigl(u_\mathrm{T} - x_i \bigr), \label{eq:chi_def}
\end{align}
\end{subequations}
for $p \in \mathbb{N}_0$. In particular, define the lifted state of zone $i$ as
\begin{equation}\label{eq:zi}
 z_i := \psi_i(\xi_i) =  
 [\Omega_{i, p}]_{p=0}^\infty,
\end{equation}
where $\Omega_{i, p} = \begin{bmatrix}
        \phi_{i,p} & \theta_{i,p} & \chi_{i,p}
    \end{bmatrix}^\top$. Note that state $x_i$, mass flow rate $\dot m_i$, input $u_\mathrm{T}$, and constant 1,  are part of $z_i$ since 
    \begin{subequations}
        \begin{align}
            &\theta_{i,1} = x_i,  \\ &\phi_{i, 0}=e^{\mu_i} = \dot m_i, \\ &\chi_{i,1}+\theta_{i,1} = u_\mathrm{T}, \\ &\phi_{i,1} =1. 
        \end{align}
    \end{subequations}

\begin{remark}
The embedding is infinite-dimensional. An exact finite-dimensional Koopman model valid for all inputs and disturbances requires the observable subspace, say $\mathcal{W}$, to be invariant under each operator in $\mathcal{K}_i$ (recall~\eqref{eq:Koopman_generator}), in
particular, under the drift $f_{0,i}\,\partial_{x_i}$ and $\partial_{\mu_i}$~\cite{goswami2017global, 11720155}.
Since $f_{0,i}\,\partial_{x_i} x_i=f_{0,i}$, we require $f_{0,i}\in \mathcal{W}$ and also $\partial_{\mu_i} f_{0,i}\in \mathcal{W}$. However, from~\eqref{eq:g_exponential},~$f_{0,i}$ depends on $\dot m_i\, e^{-c_i/\dot m_i}=\sum_{n\ge0}\alpha_{i,n}\,\dot m_i^{\,1-n}$, which has infinitely many nonzero components in eigenbasis
$\dot m_i^{\,r}=e^{r\mu_i}$, $r\in\mathbb Z$, of~$\partial_{\mu_i}$.  Hence, no finite-dimensional \(\partial_{\mu_i}\)-invariant subspace can contain $f_{0,i}$, and therefore no finite-dimensional $\mathcal{W}$ containing $x_i$ can be invariant under~$\mathcal{K}_i$.
\end{remark}

The following two subsections motivate the choice of observables~\eqref{eq:observables}.

\subsection{Linear emitter heat output in the lifted space}
Consider~\eqref{eq:rad_dynamics}, which is associated with function $g_i(x_i, u)$ in~\eqref{eq:nl_dynamics}.         Using the series computation of the exponential and substituting~$\dot m_i = e^{\mu_i}$, we have that
\begin{equation}\label{eq:exponential}
    e^{-c_i/\dot m_i} = \sum_{n=0}^\infty \dfrac{(-c_i)^n}{n!} \dfrac{1}{\dot{m}_i^n} = \sum_{n=0}^\infty \alpha_{i, n} e^{-n \mu_i}
    \end{equation}
with $\alpha_{i, n} =(-c_i)^n/n!$. Therefore, 
        \begin{equation}\label{eq:g_exponential}
\begin{split}
        g_i(\xi_i) &= e^{\mu_i}  (u_\mathrm{T}-x_i) \left(1- \sum_{n=0}^\infty \alpha_{i, n} e^{-n \mu_i}\right) \\ &=- e^{\mu_i}  (u_\mathrm{T}-x_i) \sum_{n=1}^\infty \alpha_{i, n} e^{-n \mu_i} \\ &=-    \sum_{n=1}^\infty \alpha_{i, n} e^{(1-n) \mu_i} (u_\mathrm{T}-x_i).
        \end{split}
\end{equation}
Finally, considering~\eqref{eq:chi_def}, we obtain
\begin{equation}\label{eq:Q_chi}
\begin{split}
   g_i(\xi_i) &= - \sum_{n=1}^\infty \alpha_{i, n} \chi_{i,n},
        \end{split}
\end{equation}
where, by definition, $\chi_{i,n}$ is part of the lifted state, and $\alpha_{i, n}$ are constants for $n =1, 2, ..., \infty$. Therefore, the nonlinear part of~\eqref{eq:nl_dynamics} can be written as a linear function in the lifted space, i.e., 
\begin{equation}\label{eq:xi_chi}
    \dot x_i = a_i x_i - b_i\sum_{n=1}^\infty \alpha_{i, n} \chi_{i,n} + w_i.
\end{equation}
The same  applies to the heat output of emitter $i$, which is given by~$Q^{\mathrm{rad}}_i = \cp g_i(\xi_i)$.

\subsection{Shift rule and time derivatives}
The following lemma introduces a key identity: products of the chosen observables~\eqref{eq:observables} stay in the same family, just with shifted index.

\begin{lemma}[Product closure]\label{lem:shift}
For any $p,n \in \mathbb{N}_0$, the following holds:
\begin{subequations}
\begin{align}
&\phi_{i,p}\,\phi_{i,n} = \phi_{i,\,p+n-1}, \label{eq:phiphi}\\
&\phi_{i,p}\,\theta_{i,n} = \theta_{i,\,p+n-1}, \\
&\phi_{i,p}\,\chi_{i,n} = \chi_{i,\,p+n-1}.
\end{align}
\end{subequations}
\end{lemma}

\begin{proof} Directly from the definitions of  $ \phi_{i,p}$, $\theta_{i,n}$, and~$\chi_{i,n}$, we have that
\begin{equation*}
\begin{split}
\phi_{i,p}  \phi_{i,n} &= e^{(1-p)\mu_i} e^{(1-n) \mu_i}   = e^{(2-p-n)\mu_i}   \\ &= \phi_{i,p + n -1}, \\[6pt]
    \phi_{i,p} \theta_{i,n} &= e^{(1-p)\mu_i} e^{(1-n) \mu_i} x_i \\ &= e^{(2-p-n)\mu_i}  x_i \\ &= \theta_{i,p + n -1}, \\[6pt]
    \phi_{i,p} \chi_{i,n} &= e^{(1-p)\mu_i} e^{(1-n) \mu_i} (u_\mathrm{T}-x_i)  \\ &= e^{(2-p-n)\mu_i}  (u_\mathrm{T}-x_i) \\&= \chi_{i,p + n -1},
    \end{split}
\end{equation*}
for all $p,n \in \mathbb{N}_0$. 
\end{proof}

\vspace{4pt}
\noindent The first time derivatives of \eqref{eq:observables} satisfy for any $p\in \mathbb{N}_0$,
\begin{subequations}\label{eq:time_derivatives}
\begin{align}
\dot\phi_{i,p} &= \partial_{\mu_i} \phi_{i,p} \tilde u_{\mu, i} = (1-p) \phi_{i,p} \tilde u_{\mu, i}, \hspace{2cm} \label{eq:dphi}  \\[6pt]
    \dot\theta_{i,p} &= \partial_{\mu_i}\theta_{i,p} \tilde u_{\mu, i} + \partial_{x_i} \theta_{i,p} \dot x_i  \label{eq:dtheta}  \\
     &= \partial_{\mu_i}\theta_{i,p} \tilde u_{\mu, i} + f_{0,i} \partial_{x_i} \theta_{i,p}  +  \phi_{i,p} w_i,   \hspace{1.3cm} \nonumber 
 \\[6pt]
        \dot\chi_{i,p} &=  \partial_{\mu_i} \chi_{i,p} \tilde u_{\mu, i} + \partial_{x_i} \chi_{i,p} \dot x_{i} + \partial_{u_\mathrm{T}} \chi_{i,p} \tilde u_{\mathrm{T}} \label{eq:dchi}\\&=  (1-p) \chi_{i,p} \tilde u_{\mu, i} + f_{0,i} \partial_{x_i} \chi_{i,p}  - \phi_{i,p} w_i  + \phi_{i,p}\tilde u_\mathrm{T}, \nonumber
 \end{align}
 \end{subequations}
 where
 \begin{subequations}
 \begin{equation}\label{eq:L0theta}
          \begin{split}
        f_{0,i}  \partial_{x_i} \theta_{i,p}  &=   a_i \phi_{i,p} x_i - b_i \sum_{n=1}^\infty \alpha_{i, n}  \phi_{i,p} \chi_{i,n} \\&=  a_i \theta_{i,p} - b_i \sum_{n=1}^\infty \alpha_{i, n} \chi_{i,p+n-1},
     \end{split}
 \end{equation}
      \begin{equation}\label{eq:L0chi}
          \begin{split}
        f_{0,i} \partial_{x_i} \chi_{i,p}  &= - a_i \phi_{i,p} x_i - b_i \sum_{n=1}^\infty \alpha_{i, n} \phi_{i,p} \chi_{i,n} \\&= - a_i \theta_{i,p} + b_i \sum_{n=1}^\infty \alpha_{i, n} \chi_{i,p+n-1}.
     \end{split}
     \end{equation}
 \end{subequations}
Note that~\eqref{eq:dphi} is obtained by direct differentiation, whereas for~\eqref{eq:dtheta} and \eqref{eq:dchi}, we have also used~\eqref{eq:Q_chi} and Lemma~\ref{lem:shift}.

\subsection{Assembled bilinear system}
The aggregation of time derivatives \eqref{eq:time_derivatives} in matrix form results in
\begin{equation}\label{eq:matrix_omega}
\begin{split}
    \begin{bmatrix}
    \dot\phi_{i,p} \\
    \dot\theta_{i,p} \\
        \dot\chi_{i,p}
    \end{bmatrix} = &  \begin{bmatrix}
      0 & 0  & 0 \\ 0 & a_i & -b_i \alpha_{i, 1} \\ 0 & -a_i & b_i \alpha_{i, 1}
    \end{bmatrix}  \begin{bmatrix}
    \phi_{i,p} \\
    \theta_{i,p} \\
    \chi_{i,p}
    \end{bmatrix} + \begin{bmatrix}
     0 \\ 1 \\ -1
    \end{bmatrix}  
     \phi_{i,p} w_i   \\+  &\begin{bmatrix}
     0 & 1-p & 0 & 0 \\ 0  & 0  & 1-p & 0 \\ 1 & 0 & 0 & 1-p
    \end{bmatrix}  \begin{bmatrix}
     \phi_{i,p} \tilde u_\mathrm{T} \\
      \phi_{i,p} \tilde u_{\mu, i} \\
     \theta_{i,p}  \tilde u_{\mu, i} \\
     \chi_{i,p} \tilde u_{\mu, i} 
    \end{bmatrix}
    \\+ &\begin{bmatrix}
        0 \\ -b_i \sum_{n=2}^{\infty} \alpha_{i,n}\,\chi_{i,p+n-1}  \\ b_i \sum_{n=2}^{\infty} \alpha_{i,n}\,\chi_{i,p+n-1} 
    \end{bmatrix}.
    \end{split}
\end{equation}
From the above, we observe the following structural properties for all $i\in \{1, 2, ..., n_\mathrm{z}\}$:
\begin{itemize}
    \item $\dot\phi_{i,p}$ contains bilinear term $\phi_{i,p}\tilde u_{\mu, i}$;
    \item $\dot\theta_{i,p}$ contains bilinear term $\theta_{i,p}\tilde u_{\mu,i}$ and depends linearly on observables $\theta_{i,p}$, $[\chi_{i,p+n-1}]_{n=1}^{\infty}$, and~$\phi_{i,p}$;
    \item $\dot\chi_{i,p}$ contains bilinear terms $\chi_{i,p}\tilde u_{\mu,i}$ and $\phi_{i,p}\tilde u_{\mathrm{T}}$, and depends linearly on observables $\theta_{i,p}$, $[\chi_{i,p+n-1}]_{n=1}^{\infty}$, and $\phi_{i,p}$.
\end{itemize}
 As a result, the global lifted state $z := [z_i]_{i=1}^{n_\mathrm{z}}$ evolves according to a bilinear model of the form
 \begin{equation}\label{eq:final_z_tilde}
     \dot{z} =  A z  + E (z \otimes \tilde u) + F (z \otimes w),
 \end{equation}
 where the input vector is defined as in~\eqref{eq:tilde_u} and $w=[w_i]_{i=1}^{n_\mathrm{z}}$. Note that $B$ and $G$ in \eqref{eq:lifted_system_cont} result in the zero matrices, whereas $A$, $E$, and $F$,  are defined according to~\eqref{eq:matrix_omega}.

\vspace{4pt}

\begin{proposition}[Invariance]\label{prop:inv}
\label{prop:invariance}
Consider the topological span of the observables, i.e., ${\mathcal{S} := \overline{\mathrm{span}}\{\phi_{i,p}, \theta_{i,p}, \chi_{i,p}\}_{i,p \in \mathbb{N}_0}}$. Then, $\mathcal{S}$ is invariant under input generators $\partial_{\mu_i}$, $\partial_{u_T}$, $\partial_{x_i}$, and under drift generator $f_{0, i}\partial_{x_i}$, in the closed-span~sense.
\end{proposition}

\begin{proof}
The input generators act diagonally or by a shift on the dictionary elements as shown in~\eqref{eq:dphi}--\eqref{eq:dchi}. Specifically, 
\begin{align}
\partial_{\mu_i} \phi_{i,p} &= (1-p)\phi_{i,p}, & \partial_{\mu_i} \theta_{i,p} &= (1-p)\theta_{i,p}, \notag\\
\partial_{\mu_i} \chi_{i,p} &= (1-p)\chi_{i,p},  &\partial_{u_T} \chi_{i,p} &= \phi_{i,p},   \notag\\
 \partial_{x_i} \chi_{i,p} &= -\phi_{i,p},  &
\partial_{x_i} \theta_{i,p} &= \phi_{i,p}, \notag
\end{align}
and all other combinations vanish. Each output lies in $\mathcal{S}$, so invariance under $\partial_{\mu_i}$, $\partial_{u_T}$, $\partial_{x_i}$ holds by inspection.

As for the drift generator $f_{0,i}\partial_{x_i}$, since $\theta_{i,p}$ and $\chi_{i,p}$ depend on $x_i$ only through $x_i$ and $u_T - x_i$, the chain rule gives
\[
f_{0,i} \partial_{x_i} \theta_{i,p} = f_{0,i} \phi_{i,p}, \qquad f_{0,i} \partial_{x_i} \chi_{i,p} = - f_{0,i} \phi_{i,p}.
\]
By the product closure of Lemma~1, $\phi_{i,p}x_i = \theta_{i,p}$ and $\phi_{i,p}\chi_{i,n} = \chi_{i,p+n-1}$. Substituting, we obtain~\eqref{eq:L0theta} and~\eqref{eq:L0chi}. Note that $u_\mathrm{T} - x_i$ is physically bounded, hence we can find a number $\Delta_{\max} \in \mathbb{R}$ such that 
\begin{equation}\label{eq:Delta_max}
    |u_\mathrm{T}(t) - x_i(t)| \le \Delta_{\max}.
\end{equation}
Considering this, the series in \eqref{eq:L0theta}--\eqref{eq:L0chi} converge absolutely since 
\begin{equation}
\begin{split}
  \sum_{n= 1}^\infty |\alpha_{i,n}|\,|\chi_{i,p+n-1}| &\;\le\; \Delta_{\max}\,\dot m_i^{2-p} \sum_{n=1}^\infty \frac{(c_i/\dot m_i)^n}{n!} \\&\;\le\; \Delta_{\max}\,\dot m_i^{2-p}\, e^{c_i/\dot m_i} \;<\; \infty,   
\end{split}
\label{eq:convergence}
\end{equation}
for $\dot m_i \ge \dot m_{\min,i} > 0$. Hence each right-hand side of \eqref{eq:L0theta}--\eqref{eq:L0chi} is a well-defined element of $\mathcal{S}$ in the closed-span sense, so $f_{0, i} \partial_{x_i} \theta_{i,p} \in \mathcal{S}$ and  $f_{0, i} \partial_{x_i} \chi_{i,p} \in \mathcal{S}$. Combining the two parts, $\mathcal{S}$ is invariant under all four generators. 
\end{proof}

\vspace{4pt}

Proposition~\ref{prop:inv} establishes that the span of the dictionary is invariant under the Koopman generator~\eqref{eq:Koopman_generator}.
A direct consequence is that trajectories of the nonlinear dynamics~\eqref{eq:nl_dynamics} map
to trajectories of the lifted model, as stated next.
\vspace{4pt}
\begin{corollary}\label{cor:forward}
Let $\tilde u$
be an input trajectory, and let the initial condition $z(0)$ be
determined by the initial zone temperatures
$[x_i(0)]_{i=1}^{n_\mathrm{z}}$, the initial supply temperature
$u_\mathrm{T}(0)$, and the initial mass flow rates
$[\dot m_i(0)]_{i=1}^{n_\mathrm{z}}$. If
$\xi_i(t) = (x_i(t), \mu_i(t), u_\mathrm{T}(t))$,
$i \in \{1,2,\dots,n_\mathrm{z}\}$, solves~\eqref{eq:augmented_system}, then
$z(t) = [\psi_i(\xi_i(t))]_{i=1}^{n_\mathrm{z}}$
solves~\eqref{eq:final_z_tilde}.
\end{corollary}
\begin{proof}
By Proposition~\ref{prop:inv}, $\mathcal{K}_i \psi_i$ lies in the span
of $\psi_i$ for each $i$, with coefficients given by the matrices
in~\eqref{eq:final_z_tilde}. Since
$\frac{d}{dt}\psi_i(\xi_i(t)) = (\mathcal{K}_i \psi_i)(\xi_i(t))$ along
solutions of~\eqref{eq:augmented_system}, the lifted trajectory $z(t)$
satisfies~\eqref{eq:final_z_tilde}. 
\end{proof}
\vspace{4pt}

As for the reverse direction, Proposition~\ref{prop:equivalence} in Appendix~\ref{sec:equivalence_trajectories} shows that every solution of
\eqref{eq:final_z_tilde} starting from $\psi(\xi(0))$, within a suitable class
of solutions, is the lift of a solution of~\eqref{eq:nl_dynamics}. 

\subsection{Finite-dimensional truncation and error bound}

For implementation, we truncate $z_i$ in~\eqref{eq:zi} at a finite order $\bar P \in \mathbb{N}$, retaining only $\Omega_{i,p}$ for $p = 0,1,\dots,\bar P$, i.e.,
\begin{equation}\label{eq:zi_trunc}
    z_i^{(\bar P)} \coloneqq [\Omega_{i,p}]_{p=0}^{\bar P} \in \mathbb{R}^{3(\bar P+1)}.
\end{equation}
\noindent This truncation has a direct consequence on the dynamics~(see~\eqref{eq:matrix_omega}): for each retained index $p \in \{0,1,\dots,\bar P\}$, the sum $\sum_{n=1}^{\infty} \alpha_{i,n}\,\chi_{i,p+n-1}$ involves observables $\chi_{i,p+n-1}$, which remain within the truncated dictionary~\eqref{eq:zi_trunc} only while $p+n-1 \le \bar P$, i.e., $n \le \bar P - p + 1$.  Consequently, the truncated dictionary~\eqref{eq:zi_trunc} is not invariant under Koopman generator~\eqref{eq:Koopman_generator}.

Setting $N(p) \coloneqq \bar P - p + 1$, we can decompose
\begin{equation}
\begin{split}
        &\sum_{n=1}^{\infty} \alpha_{i,n}\,\chi_{i,p+n-1} = \sum_{n=1}^{N(p)} \alpha_{i,n}\,\chi_{i,p+n-1} + r_{i,p}^{(N(p))},
\end{split}
\end{equation}
where
\begin{equation}\label{eq:remainder}
  r_{i,p}^{(N(p))} =  \sum_{n=N(p)+1}^{\infty} \alpha_{i,n}\,\chi_{i,p+n-1}
\end{equation}
is the remainder that is neglected. This term determines the truncation error incurred by the finite dictionary~\eqref{eq:zi_trunc}, and can be bounded as shown next.

\begin{proposition}[Truncation error bound]\label{prop:trunc_bound}
Let \mbox{$\dot m_{\min ,i} > 0$} and $\dot m_i(t) \in [\dot m_{\min ,i}, \dot m_{\max ,i}]$,  $\mathrm{NTU}_{\max,i} = c_i / \dot m_{\min,i}$, and $\bar M_p \coloneqq \max\{\dot m_{\min,i}^{\,2-p},\, \dot m_{\max,i}^{\,2-p}\}$. Then, for any $p \in \mathbb{N}_{0}$ and truncation order $\bar P \in \mathbb{N}$, the truncation remainder $r_{i,p}^{(N(p))}$ satisfies
\begin{equation}\label{eq:trunc_bound}
\begin{aligned}
    \bigl|r_{i,p}^{(N(p))}\bigr| \;& \le\; \Delta_{\max}\, \bar M_p \, \frac{\mathrm{NTU}_{\max, i}^{\,N(p)+1}}{(N(p)+1)!}\, e^{\mathrm{NTU}_{\max, i}}. 
\end{aligned}
\end{equation}
For $p \geq 2$, the right-hand side of \eqref{eq:trunc_bound} can be rewritten as 
\begin{align}
    \bigl|r_{i,p}^{(N(p))}\bigr| \leq \Delta_{\max} \dot m_{\mathrm{min},i}^{-\bar P}  \frac{c_i^{\,\bar P - p + 2}}{(N(p)+1)!}\,  e^{\mathrm{NTU}_{\max, i}}.   \nonumber
\end{align}
\end{proposition}

\vspace{5pt}
\noindent The proof of Proposition~\ref{prop:trunc_bound} is presented in Appendix~\ref{app:proofs}. 
\vspace{2pt}

As shown in~\eqref{eq:trunc_bound}, for a fixed $p$, the error decays factorially in $N(p)+1$. For a fixed $\bar P$, the bound is largest at $p=\bar P$, where $N(p) = 1$, so the factorial decay does not hold uniformly over the retained observables.
More importantly, the error at any $p$ vanishes as $\bar P \to \infty$, confirming that the truncated model~\eqref{eq:zi_trunc} recovers the exact bilinear representation in the limit ${\bar P \to \infty}$. Complementing this result, the following proposition propagates bound~\eqref{eq:trunc_bound} along system trajectories to bound the error in the zones temperature. 

\begin{proposition}[Trajectory bound]\label{prop:trunc_tray}
Let $\bar P$ be the truncation order and $x_i^{(\bar P)}$ denote the corresponding
zone-temperature prediction for zone $i\in \{1, 2, ..., n_\mathrm{z}\}$. Define
\begin{equation}
    \begin{split}
        L_i &\coloneqq |a_i| + b_i\,\dot{m}_{\max,i}e^{\mathrm{NTU}_{\max, i}},
\\
\bar{\tau}_i &\coloneqq \sup_{t\ge0,\,p\le \bar{P}}\ \dot{m}_i(t)^{p-1}
\bigl|r_{i,p}^{(N(p))}(t)\bigr|,
    \end{split}
\end{equation}
where $\bar\tau_i<\infty$ by Proposition~\ref{prop:trunc_bound}. Then, for all $t\ge0$,
\[
\bigl|x_i(t)-x_i^{(\bar P)}(t)\bigr| \le \frac{b_i\bar{\tau}_i}{L_i}\bigl(e^{L_it}-1\bigr).
\]
\end{proposition}

\vspace{5pt}
\noindent The proof of Proposition~\ref{prop:trunc_tray} is also  given in Appendix~\ref{app:proofs}.

\begin{remark}
    Accurately identifying the parameters of building thermal models~\eqref{eq:zone_base} and~\eqref{eq:wall_temp} is challenging due to unknown thermal disturbances and parameter identifiability issues~\cite{7551202}. In hydronic heating systems, this can be further complicated by the nonlinearities. Another interesting advantage of reformulating the original nonlinear system~\eqref{eq:ss_T} as bilinear model~\eqref{eq:final_z_tilde} is that, when some system parameters are unknown, 
    which is often the case in practice, the matrices in~\eqref{eq:final_z_tilde} can be identified from measured data through a linear regression problem. In contrast, identifying the original nonlinear model is generally more challenging, since the resulting problem is nonconvex and can therefore converge to a local solution rather than the true model.
\end{remark}

\section{Controller design}\label{sec:controller_design}

This section designs an MPC  based on lifted model~\eqref{eq:final_z_tilde}, which optimizes the reparameterized inputs~$\tilde u$. The valve positions that realize the resulting flow rates are recovered afterwards.

\subsection{Hydraulic constraints}

\begin{figure}
    \centering
    \includegraphics[width=\columnwidth]{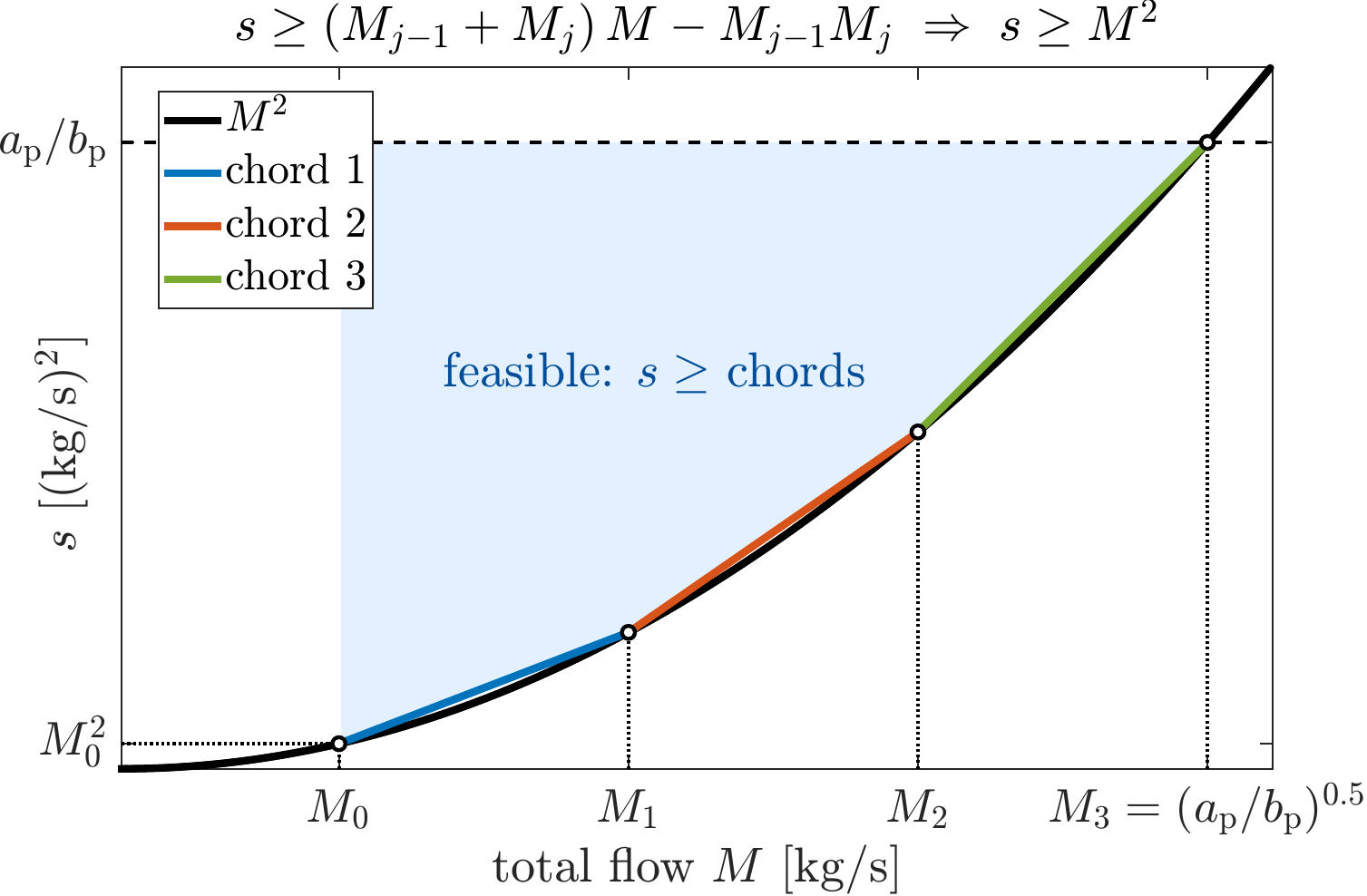}
    \caption{Schematic illustration of the approximation of $M^2$ using three chords for the constraint~\eqref{eq:pump}.}
    \label{fig:scheme_s_variable}
\end{figure}

According to the hydraulic model in Section~\ref{sec:hydraulic_model}, flow rates $\dot m_i$, $i \in \{1,2,\dots,n_\mathrm{z}\}$, can be realized by valve positions $u_{\mathrm{v},i} \in (0,1]$ if and only~if
\begin{equation}
  \kappa_{\min,i}\, \dot m_i^2 + \rho^\mathrm{wat} g b_\mathrm{p} M^2 \le \rho^\mathrm{wat} g a_\mathrm{p}, \quad i \in \{1,2,\dots,n_\mathrm{z}\},
  \label{eq:pump}
\end{equation}
where $\kappa_{\min,i}$ denotes the minimum hydraulic resistance of radiator $i$, corresponding to a fully open valve. Considering the observables, let us define $\dot m_i^2 = \phi_{i, -1} = e^{2\mu_i}$ and $M = \sum_{i} \phi_{i,0}$, where $\phi_{i, -1} $ evolves according to
\begin{equation}
    \dot \phi_{i, -1}  = 2\phi_{i, -1}  \tilde u_{\mu, i}. \label{eq:phi_flow_dynamics}
\end{equation}
Then, the first term of inequality~\eqref{eq:pump} becomes linear in~$\phi_{i,-1}$, which can be incorporated into the model as a new observable with bilinear dynamics.
The term with $M^2$, however, contains products $\dot m_i \dot m_j$ of different zones, which cannot be written as a function of the lifted states of a single zone. Therefore, $M^2$ is replaced by a new decision variable~$s$, and~\eqref{eq:pump} becomes
\begin{align}
  \kappa_{\min,i}\,\phi_{i,-1} + \rho^\mathrm{wat} g b_\mathrm{p}\, s &\le \rho^\mathrm{wat}  g a_\mathrm{p}, \label{eq:pump_lift}\\
  (M_{j-1}+M_j)\,M - M_{j-1}M_j &\le s, \label{eq:chords}
\end{align}
where $j \in \{ 1,\dots,n_\mathrm{s} \}$, $n_\mathrm{s} \in \mathbb{N}_{>1}$. Here, $M_0 < M_1 < \dots < M_{n_\mathrm{s}}$ are fixed breakpoints, with
$M_0 = \sum_{i=1}^{n_\mathrm{z}} \dot m_{\min,i}$ the total minimum flow and
$M_{n_\mathrm{s}} = (a_\mathrm{p}/b_\mathrm{p})^{1/2}$ the maximum (see Figure~\ref{fig:scheme_s_variable}). Each inequality \eqref{eq:chords} is the
chord of $M^2$ through points $(M_{j-1}, M_{j-1}^2)$ and $(M_j, M_j^2)$. Since $M^2$
is convex, each chord lies above $M^2$ on its own segment $[M_{j-1}, M_j]$ and below it
elsewhere. Hence, for any $M \in [M_0, M_{n_\mathrm{s}}]$, the largest left-hand side in
\eqref{eq:chords} is attained by the segment containing $M$, so that
\eqref{eq:chords} enforces $s \ge M^2$. 
Consequently, every flow trajectory satisfying \eqref{eq:pump_lift}--\eqref{eq:chords}
also satisfies \eqref{eq:pump}, and all constraints remain linear in the lifted~space.

\subsection{MPC problem} 

For use in MPC, lifted model~\eqref{eq:final_z_tilde} must be discretized. To this end, one can use the exact method described in Appendix~\ref{app:disc}, assuming that the inputs are held constant over each sampling interval, although the resulting discrete-time model loses its bilinear structure. Alternatively, forward Euler discretization can be used. To accommodate different discretization methods, which are compared in Section~\ref{sec:NCS}, 
let us write them in the general form~as
\begin{equation}\label{eq:disc_ztilde}
    z(k+1) = f_\mathrm{d}\big(z(k), \tilde u(k),w(k)\big),
\end{equation}
where $f_\mathrm{d}$ denotes the discrete-time dynamics and $k \in \mathbb{N}_0$ is the time index.

The objective is to maintain every zone $i\in\{1,2,\ldots,n_\mathrm{z}\}$ within the desired comfort temperature range while reducing energy use. Considering this, at every time step  $k \in \mathbb{N}_0$, we solve the following MPC problem for the lifted system:
\begin{subequations}\label{eq:MPC}
\begin{align*}
   \min_{\boldsymbol{\delta}(k), \boldsymbol{\tilde u}(k), \boldsymbol{s}(k)} \
   \sum_{i=1}^{n_\mathrm{z}} &\sum_{t=1}^{N_\mathrm{p}} \Big( \lambda_i(k+t)\, \delta_{i}(t|k)^2 + \rho_Q t_\mathrm{s}\, Q^{(\bar{P})}_i(t|k) \Big)
\end{align*}
\begin{align}
& \text{s.t.} \hspace{7.5cm}\nonumber \\
    & z(0|k) = [z_{i}(0|k)]_{i=1}^{n_\mathrm{z}}, \label{eq:init_cond} \\
    & z(t+1|k) = f_\mathrm{d}\big(z(t|k), \tilde u(t|k),w(t|k)\big), \\
    & \theta_{i,1}(t+1|k) \in [\underline T_i(t+1|k), \ \overline T_i(t+1|k)], \\
    & \underline T_i(t+1|k) = T_{\min, i}(k+t+1) - \delta_{i}(t+1|k), \\
    & \overline T_i(t+1|k) = T_{\max, i}(k+t+1) + \delta_{i}(t+1|k), \\
    & \delta_{i}(t+1|k) \ge 0, \label{eq:slack_pos} \\
    & \phi_{i,0}(t+1|k) \in [\dot m_{\min,i}, \  \dot m_{\max,i}], \label{eq:flow_floor} \\
    & u_\mathrm{T}(t+1|k) = u_\mathrm{T}(k) + t_\mathrm{s} \sum_{\tau=0}^{t} \tilde u_\mathrm{T}(\tau|k), \\
    &u_\mathrm{T}(t+1|k) \in [T^{\mathrm{sup}}_{\min}, \ T^{\mathrm{sup}}_{\max}], \label{eq:Tsup} \\[4pt]
    & \kappa_{\min,i}\, \phi_{i,-1}(t+1|k) + \rho^\mathrm{wat} g b_\mathrm{p}\, s(t+1|k) \le \rho^\mathrm{wat} g a_\mathrm{p}, \\
& \phi_{i,-1}(0|k) = \dot m_i(k)^2, \\
&  \phi_{i,-1} (t+1|k)  = \bigl(1 + 2 t_\mathrm{s} \tilde u_{\mu,i}(t|k)\bigr)\,\phi_{i,-1}(t|k), \label{eq:phi_discrete}\\
    & s(t+1|k) \ge (M_{j-1} + M_j)\, M(t+1|k) - M_{j-1} M_j, \nonumber \\
    & \qquad \qquad j \in \{1,2, \dots, n_\mathrm{s}\}, \\
    & M(t+1|k) = \sum_{i=1}^{n_\mathrm{z}} \phi_{i,0}(t+1|k), \\
    & 0 \le s(t+1|k) \le a_\mathrm{p}/b_\mathrm{p}, \\
    & i\in \{1, 2, \dots, n_\mathrm{z}\}, \ t \in \{0,1, \dots, N_{\mathrm{p}}-1\},
\end{align}
\end{subequations}
where $t_\mathrm{s} \in \mathbb{R}_{>0}$ is the sampling time, $N_{\mathrm{p}} \in \mathbb{N}$ is the prediction horizon, $(t|k)$ indicates a prediction for time step $k+t$ computed at instant $k$, $\lambda_i$ are weighting factors for $i\in \{1, 2, ..., n_\mathrm{z}\}$, and $\rho_Q$ is the weighting factor of the energy use. Moreover, $Q^{(\bar{P})}_i(t|k)$ is the approximated heat output of radiator $i$ as
\begin{equation}
  Q^{(\bar{P})}_i(t|k) \coloneqq -\cp \sum_{n=1}^{\bar{P}} \alpha_{i,n}\, \chi_{i,n}(t|k).
  \label{eq:Q_lift}
\end{equation}
Additionally, 
\begin{align}
    &\boldsymbol{\tilde u}\!=\!\begin{bmatrix} \tilde u(0|k) \\ \tilde u(1|k)\\ \vdots \\ \tilde u(N_{\mathrm{p}}\!-\!1|k) \end{bmatrix}\!\!,
    \  \boldsymbol{\delta}\!=\!\begin{bmatrix} \delta(1|k) \\ \delta(2|k) \\ \vdots  \\ \delta(N_{\mathrm{p}}|k) \end{bmatrix}\!\!,
    \  \boldsymbol{s}\!=\!\begin{bmatrix} s(1|k) \\ s(2|k) \\ \vdots  \\ s(N_{\mathrm{p}}|k) \end{bmatrix}\!\!,\nonumber
\end{align}
are the sequences to be optimized, with $\delta(t|k) = [\delta_{i}(t|k)]_{i=1}^{n_\mathrm{z}}$ and $s(t|k)$ an upper bound of $M(t|k)^2$, approximated by the chords in Figure~\ref{fig:scheme_s_variable}. Moreover, $T_{\min, i}(k)$ and $T_{\max, i}(k)$ are the user-specified minimum and maximum comfort temperatures for zone $i$ at instant $k$, respectively. Note that \eqref{eq:phi_discrete} is an approximate discretization of \eqref{eq:phi_flow_dynamics} obtained using forward Euler. It slightly underestimates the squared flow rate, so the hydraulic constraint is guaranteed for the predicted flow rates rather than the applied ones. The exact discretization $\phi_{i,-1}(t+1|k) = e^{2 t_\mathrm{s}\tilde u_{\mu,i}(t|k)}\,\phi_{i,-1}(t|k)$ could be used instead to recover this guarantee.

The first element of the resulting optimal sequences determines the inputs applied to the plant at time $k$. Specifically, the supply temperature and the mass flow rates are updated~as
\begin{align}
    u_\mathrm{T}(k+1) &= u_\mathrm{T}(k) + t_\mathrm{s}\tilde u_\mathrm{T}^\ast(0|k), \nonumber \\
    \dot m_i(k+1) &= \dot m_i(k)\, e^{t_\mathrm{s}\tilde u_{\mu,i}^\ast(0|k)}, \nonumber
\end{align}
where $^\ast$ refers to optimal. The valve positions $[u_{\mathrm{v}, i}(k+1)]_{i=1}^{n_{\mathrm{z}}}$ that realize these flows follow from the pump curve as
\begin{equation}
\begin{split}
 u_{\mathrm{v}, i}(k+1) = \dfrac{\kappa_{\mathrm{v}, i}}{\sqrt{\kappa_i(k+1) - \kappa_{0, i}^2}}, \nonumber \\
 \kappa_i(k+1) = \dfrac{\rho^{\mathrm{wat}} g \big(a_{\mathrm{p}} - b_{\mathrm{p}} M(k+1)^2\big)}{\dot m_i(k+1)^2}, \nonumber
\end{split}
\end{equation}
with $M(k+1) = \sum_{i=1}^{n_\mathrm{z}} \dot m_i(k+1)$. Once these inputs are implemented, \eqref{eq:MPC} is solved again at the next sampling instant, following the receding-horizon approach.

\begin{remark}
  The lifted initial condition in problem~\eqref{eq:MPC} requires $\dot m_i(k)$ and $u_\mathrm{T}(k)$. While the latter is an actuator signal, the former is assumed to be measured at every time step $k \in \mathbb{N}_{\geq 0}$. Note also that constraints~\eqref{eq:const_mT}  become state constraints on $\phi_{i,0} = \dot{m}_i$ and $\theta_{i,1} + \chi_{i,1} = u_\mathrm{T}$.
\end{remark}

\section{Numerical Case Study} \label{sec:NCS}

\begin{figure}[!t]
    \centering

    \begin{subfigure}[t]{0.62\linewidth}
        \centering
        \includegraphics[width=\linewidth]{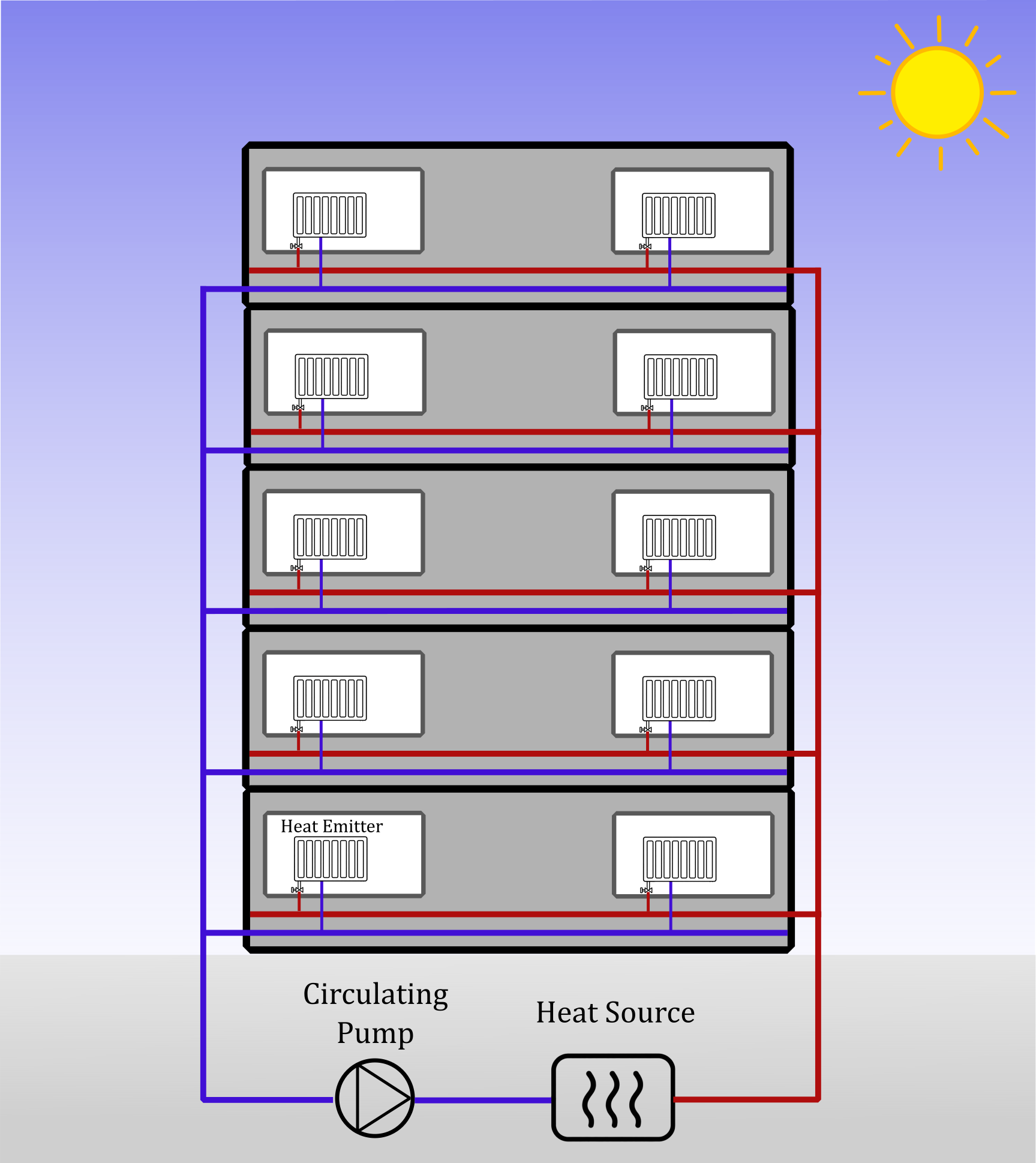}
        \caption{Schematic of the five-story case study building with heat emitters, a circulating pump, and a heat source.}
        \label{fig:NCS_building}
    \end{subfigure}

    \vspace{0.5em}

    \begin{subfigure}[t]{0.65\linewidth}
        \centering
        \includegraphics[width=\linewidth]{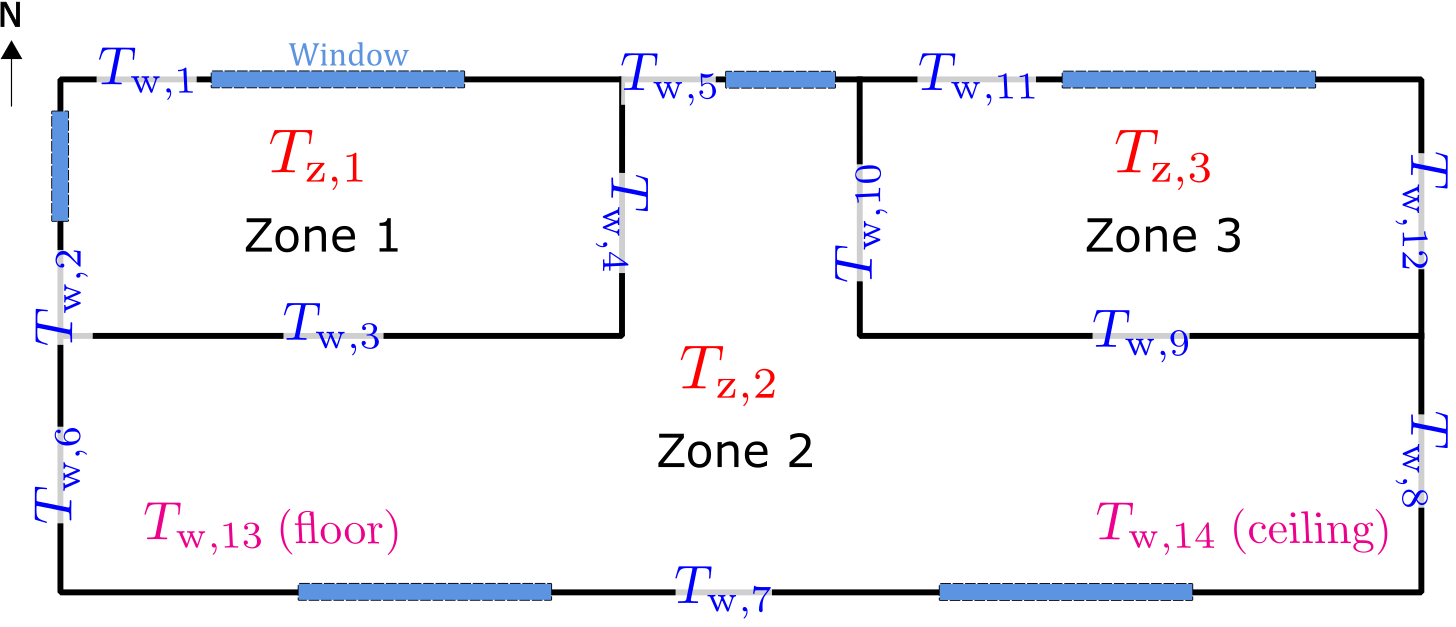}
        \caption{Floor plan of the ground floor.}
        \label{fig:NCS_floor_map}
    \end{subfigure}

    \caption{Schematic representation of the considered five-story building and an example floor plan.}
    \label{fig:NCS_case_study}
\end{figure}

We consider a numerical case study involving a five-story building with 20 zones (three zones on the ground floor, five on the first floor, and four on each of the remaining three floors), with one hydronic heat emitter installed in each zone. All heat emitters in the building are supplied by a common heat source (e.g., a boiler) through a circulating pump, as shown in Figure~\ref{fig:NCS_building}. The floor plan of the ground floor, including the zones and walls states, is shown in Figure~\ref{fig:NCS_floor_map} as an example. Note that the thermal ceiling of the ground floor also serves as the thermal floor of the first floor, and the same applies to all other adjacent~stories.  Regarding hydraulic modeling of the building, as commonly encountered in practice, the heat emitters on the higher floors have higher hydraulic resistance than those on the lower floors, represented by smaller values of $k_{0,i} \in \mathbb{R}_{>0}$ in~\eqref{eq:flow_coefficients}.

\subsection{Koopman Representation Accuracy}
\begin{figure}[!t]
    \centering
    \includegraphics[width=0.9\linewidth]{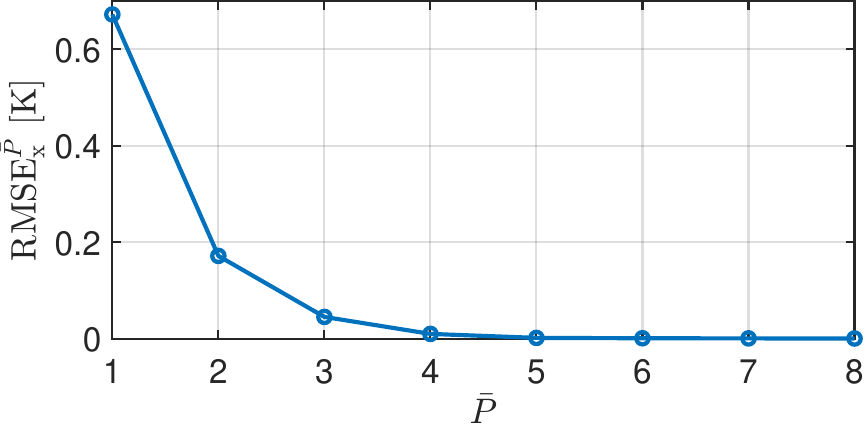}
    \caption{Zone-temperature RMSE between original nonlinear system~\eqref{eq:ss_T} and lifted system~\eqref{eq:final_z_tilde} as a function of  $\bar{P}$.}
    \label{fig:NCS_RMSE}
\end{figure}

\begin{figure}[!t]
    \centering
    \includegraphics[width=0.95\columnwidth]{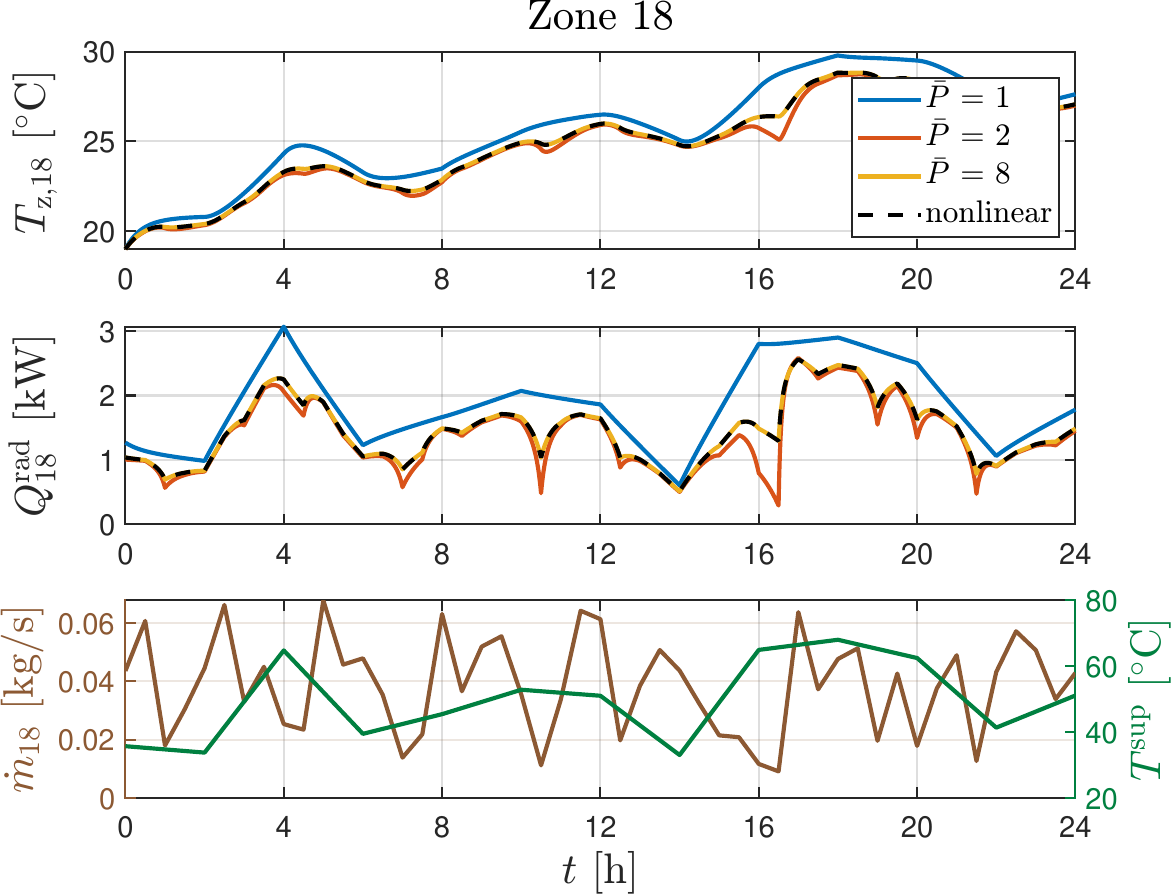}
    \caption{Temperature of Zone 18 and the heat output of its emitter obtained from nonlinear system~\eqref{eq:ss_T} and lifted system~\eqref{eq:final_z_tilde} with different values of $\bar{P}$ when applying random piecewise-linear flow-rate profiles $\dot{m}_i$, $i\in\{1, 2, ..., 20\}$, and a piecewise-linear supply temperature~$T^{\mathrm{sup}}$.}
    \label{fig:NCS_traj}
\end{figure}

This subsection assesses the accuracy of the Koopman bilinear representation~\eqref{eq:final_z_tilde} with the truncated lifted states~\eqref{eq:zi_trunc} relative to the original nonlinear system~\eqref{eq:ss_T}, and investigate how the approximation accuracy changes for different values of truncation order $\bar{P}$. To do so, we generate a piecewise-linear flow profile $\dot{m}_i$ per zone and a piecewise-linear supply temperature~$T^{\mathrm{sup}}$. They are applied to the nonlinear system~\eqref{eq:ss_T} to obtain the evolution of its states, and are also transformed into $\tilde{u}$ according to~\eqref{eq:tilde_u} and applied to lifted system~\eqref{eq:final_z_tilde}. The root mean squared error (RMSE) and maximum pointwise deviation between their zone temperatures are computed, together with the RMSE of the heat-emitter outputs. Specifically, we consider the following indicators:
\begin{align}
\mathrm{RMSE}_{\mathrm{x}}^{\bar P} &= \sqrt{\frac{1}{n_{\mathrm{z}} N_{\mathrm{t}}}\sum_{i=1}^{n_\mathrm{z}}\sum_{k=1}^{N_{\mathrm{t}}}
\bigl( C_{\mathrm{x}}z^{(\bar{P})}_i(t_k) - x_i(t_k) \bigr)^2}, \nonumber \\[15pt]
\mathrm{RMSE}_{\mathrm{Q}}^{\bar P} &= \sqrt{\frac{1}{n_{\mathrm{z}} N_{\mathrm{t}}}\sum_{i=1}^{n_{\mathrm{z}}}\sum_{k=1}^{N_{\mathrm{t}}}
\bigl( Q_i^{(\bar P)}(t_k) - Q^{\mathrm{rad}}_i(t_k) \bigr)^2}, \nonumber \\[15pt]
e_{\max}^{\bar P} &= \max_{i,k}\ \bigl| C_{\mathrm{x}}z^{(\bar{P})}_i(t_k) - x_i(t_k) \bigr|, \nonumber
\end{align}
where $n_{\mathrm{z}} = 20$ is the number of zones, $N_{\mathrm{t}} \in \mathbb{N}$ is the number of samples, $Q_i^{(\bar P)}$ is defined in \eqref{eq:Q_lift}, and 
\begin{align}
C_{\mathrm{x}}
&\coloneqq
\begin{bmatrix}
0_{1\times4}
&
1
&
0_{1\times (3\bar P - 2)}
\end{bmatrix}. \nonumber
\end{align}
The results are summarized in Table~\ref{tab:koopman_accuracy} and Figure~\ref{fig:NCS_RMSE}. Moreover, as a representative example, the temperature of Zone~18 and the heat output of its heat emitter are illustrated in Figure~\ref{fig:NCS_traj}. 

\begin{table}[!t]
\centering
\caption{Errors for different truncation
orders $\bar{P}$ of the Koopman representation.}
\label{tab:koopman_accuracy}
\begin{tabular}{cccc}
\hline
\noalign{\vskip 2pt}
$\bar{P}$ &
$\mathrm{RMSE}_{\mathrm{x}}^{\bar P}$ (K) &
$e_{\max}^{\bar P}$ (K) &
$\mathrm{RMSE}_{\mathrm{Q}}^{\bar P}$ (W) \\
\hline
1  & 0.6721 & 2.3790 & 603.5 \\
2  & 0.1714 & 1.2919 & 178.3 \\
3  & 0.0449 & 0.4859 & 54.3  \\
5  & 0.0013 & 0.0170 & 2.7   \\
8  & 0.0001 & 0.0016 & 0.1   \\
10 & 0.0000 & 0.0000 & 0.0   \\
\hline
\end{tabular}
\end{table}

\subsection{Koopman-Based MPC Performance}

To evaluate the proposed Koopman-based MPC~\eqref{eq:MPC}, we compare it with two classes of alternative formulations. The first is NMPC, which uses the original nonlinear model and serves as a benchmark. We consider two NMPC variants: one treats the wall temperatures as states, and the other, as in the Koopman-based MPC, treats them as exogenous inputs.  The second class is linearization-based MPC, which solves a quadratic program (QP). Here, we consider LTI and LTV MPC based on the original nonlinear and lifted models. In all cases, the plant is simulated with the nonlinear model~\eqref{eq:ss_T}, and the controllers are compared in terms of realized closed-loop cost and solve time. When the wall temperatures are treated as exogenous inputs (recall Assumption~\ref{ass:wall_slow}), they are predicted over the horizon via~\eqref{eq:wall_temp}, using the zone temperatures predicted at the previous time step, shifted by one step with the last value repeated.

The NMPC is formulated as
\begin{subequations}\label{eq:NMPC}
\begin{align*}
   \min_{\boldsymbol{\delta}(k), \boldsymbol{u}(k)} \
   \sum_{i=1}^{n_\mathrm{z}} &\sum_{t=1}^{N_\mathrm{p}} \Big( \lambda_i(k+t)\, \delta_{i}(t|k)^2+ \rho_Q t_\mathrm{s}\, Q_i(t|k) \Big)
\end{align*}
\begin{align}
& \text{s.t.} \hspace{7.5cm}\nonumber \\
    & T(0|k) = T(k), \label{eq:NMPC_init} \\
    & T(t+1|k) = \mathcal F\big(T(t|k), u(t|k), d(t|k)\big), \label{eq:NMPC_model} \\
    & Q(t+1|k) = \mathcal F_Q\big(T(t|k), u(t|k), d(t|k)\big), \label{eq:NMPC_heat} \\[4pt]
    & T_{\mathrm{z},i}(t+1|k) \in [\underline T_i(t+1|k), \ \overline T_i(t+1|k)], \label{eq:NMPC_comfort} \\
    & \underline T_i(t+1|k) = T_{\min ,i}(k+t+1) - \delta_{i}(t+1|k), \\
    & \overline T_i(t+1|k) = T_{\max ,i}(k+t+1) + \delta_{i}(t+1|k), \\
    & \delta_{i}(t+1|k) \ge 0, \\
    & \dot m_i(t|k) \in [\dot m_{\min,i}, \  \dot m_{\max,i}], \label{eq:NMPC_floor} \\
    & u_{\mathrm{v},i}(t|k) \in [0, \ 1], \\
    & u_\mathrm{T}(t|k) \in [T^{\mathrm{sup}}_{\min}, \ T^{\mathrm{sup}}_{\max}], \\
    & u(t|k) \coloneqq \begin{bmatrix}u_\mathrm{T}(t|k) & u_\mathrm{v}(t|k)^\top \end{bmatrix}^\top,
\end{align}
\end{subequations}
where $i \in \{1,2, \dots, n_\mathrm{z}\}$ and $t \in \{0,1, \dots, N_\mathrm{p}-1\}$. Moreover,~$\mathcal F$ is one fourth-order Runge--Kutta step of length $t_\mathrm{s}$ of~\eqref{eq:ss_T}, and $\mathcal F_Q$ gives the mean heat outputs 
\[
Q(t+1|k) \coloneqq [Q_i(t+1|k)]_{i=1}^{n_\mathrm{z}}
\]
of the heat emitters over the sampling interval from $t$ to $t+1$. 

The LTV and LTI MPCs based on the original system solve a problem similar to~\eqref{eq:NMPC}, but with every nonlinear term linearized around a nominal trajectory. For example,~\eqref{eq:NMPC_model} is linearized as
\begin{align}
T(t+1|k) &= T^{\mathrm{nom}}(t+1|k)
  + \frac{\partial \mathcal F}{\partial T} \big(T(t|k) - T^{\mathrm{nom}}(t|k)\big)
  \nonumber \\ 
  &\qquad +\frac{\partial \mathcal F}{\partial u} \big(u(t|k) - u^{\mathrm{nom}}(t|k)\big), \nonumber
\end{align}
where $T^{\mathrm{nom}}(t|k)$ is the prediction of the nonlinear model from $T(k)$ with the nominal inputs $u^{\mathrm{nom}}(t|k)$.  LTV MPC takes the optimal plan computed at $k-1$, shifted by one step, as nominal inputs and evaluates the Jacobians at every step of the nominal trajectory. By contrast, LTI MPC takes the inputs applied at $k-1$ as nominal inputs and evaluates the Jacobians once. 
To improve the performance of LTI and LTV MPCs, we additionally bound the valve moves as  
\begin{align}
    |u_{\mathrm{v},i}(t|k) - u_{\mathrm{v},i}(t-1|k)| \le \Delta u_{\mathrm{v},\max}, \label{eq:LTV_LTI_constraint}
\end{align}
$\Delta u_{\mathrm{v},\max} \in \mathbb{R}_{> 0}$, with $u_{\mathrm{v},i}(-1|k) := u_{\mathrm{v}, i}(k-1)$, thereby keeping the predicted trajectory close to the region where the linearized hydraulics are accurate. 

\begin{table}[!t]
    \centering
    \caption{Performance comparison of the MPC approaches over 10 scenarios. ``Mean max.'' denotes the average of the maximum computation times across all scenarios.}
    \label{tab:mpc_comparison}
    \renewcommand{\arraystretch}{1.12}
    \setlength{\tabcolsep}{2.5pt}
    \footnotesize
    \resizebox{\columnwidth}{!}{%
    \begin{tabular}{@{}lcccc@{}}
        \hline
        \textbf{Controller}
        & \multicolumn{2}{c}{\textbf{vs. NMPC [\%]}}
        & \multicolumn{2}{c}{\textbf{Comp. time [s]}} \\
        & \textbf{Avg.} & \textbf{Worst}
        & \textbf{Mean} & \textbf{Mean max.} \\
        \hline
        NMPC (walls as states)
        & -- & -- & 7.47 & 15.21 \\
        NMPC (walls as exog. inputs)
        & $+0.05$ & $+0.13$ & 1.94 & 3.92 \\
        \hline
        \multicolumn{5}{@{}l}{\textbf{Lifted MPC}} \\
        \multicolumn{5}{@{}l}{\quad\textit{Exact discretization}} \\
        \quad $\bar P=4$
        & $+0.02$ & $+0.35$ & 1.25 & 3.01 \\
        \quad $\bar P=5$
        & $+2.8$ & $+10.9$ & 1.38 & 4.70 \\
        \quad $\bar P=6$
        & $+0.16$ & $+0.34$ & 1.56 & 3.64 \\
        \quad $\bar P=7$
        & $-0.01$ & $+0.09$ & 1.79 & 3.86 \\
        \quad $\bar P=8$
        & $+0.08$ & $+0.15$ & 2.03 & 4.42 \\
        \multicolumn{5}{@{}l}{\quad\textit{Forward Euler (1 step)}} \\
        \quad $\bar P=4$
        & $+1.4$ & $+5.6$ & 1.11 & 2.56 \\
        \quad $\bar P=5$
        & $+7.2$ & $+23.0$ & 1.27 & 7.65 \\
        \quad $\bar P=6$
        & $+1.5$ & $+5.7$ & 1.36 & 2.97 \\
        \quad $\bar P=7$
        & $+1.4$ & $+5.7$ & 1.17 & 2.60 \\
        \quad $\bar P=8$
        & $+1.4$ & $+5.6$ & 1.60 & 3.30 \\
        \multicolumn{5}{@{}l}{\quad\textit{Forward Euler (2 substeps)}} \\
        \quad $\bar P=4$
        & $+0.26$ & $+1.0$ & 1.36 & 3.18 \\
        \quad $\bar P=5$
        & $+4.2$ & $+13.8$ & 1.53 & 6.77 \\
        \quad $\bar P=6$
        & $+0.39$ & $+1.1$ & 1.75 & 3.49 \\
        \quad $\bar P=7$
        & $+0.22$ & $+0.96$ & 2.05 & 4.20 \\
        \quad $\bar P=8$
        & $+0.31$ & $+0.99$ & 2.31 & 4.74 \\
        \hline
        \multicolumn{5}{@{}l}{\textbf{Lifted LTV MPC}} \\
        \quad $\bar P=4$
        & $+4.2$ & $+7.9$ & 0.44 & 2.88 \\
        \quad $\bar P=5$
        & $+14.0$ & $+30.1$ & 0.40 & 4.25 \\
        \quad $\bar P=6$
        & $+8.8$ & $+13.0$ & 0.51 & 5.78 \\
        \quad $\bar P=7$
        & $+6.7$ & $+12.3$ & 0.37 & 2.84 \\
        \quad $\bar P=8$
        & $+8.8$ & $+12.9$ & 0.42 & 4.61 \\
        \hline
        LTV MPC
        & $+22.0$ & $+81.9$ & 0.92 & 8.28 \\
        LTI MPC
        & $+45.8$ & $+195$ & 0.88 & 5.75 \\
        \hline
    \end{tabular}%
    }
\end{table}

\begin{figure*}[!t]
    \centering

    \begin{subfigure}[t]{0.32\textwidth}
        \centering
        \includegraphics[width=\linewidth]{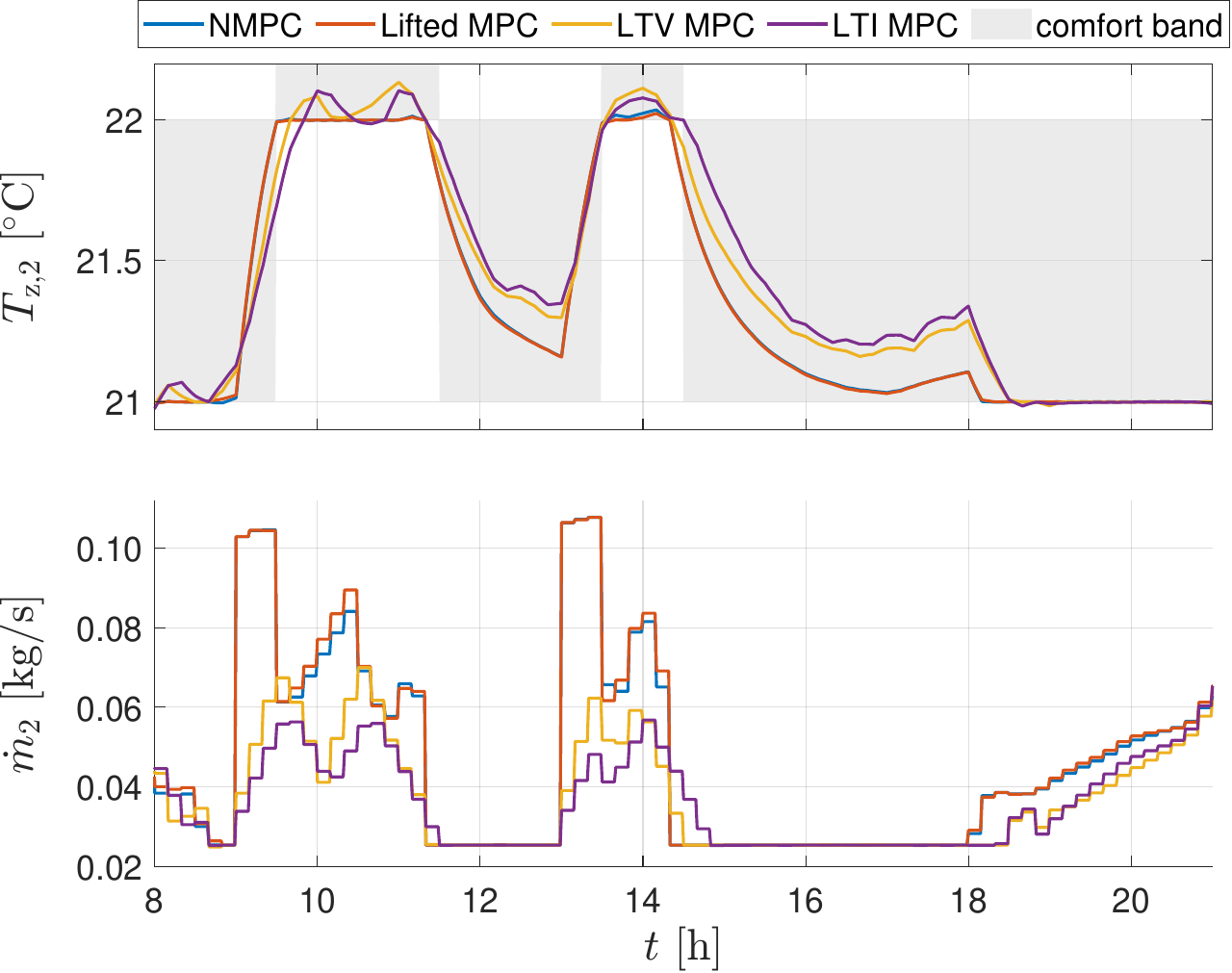}
        \caption{Zone 2, ground floor.}
        \label{fig:zone2}
    \end{subfigure}
    \hfill
    \begin{subfigure}[t]{0.32\textwidth}
        \centering
        \includegraphics[width=\linewidth]{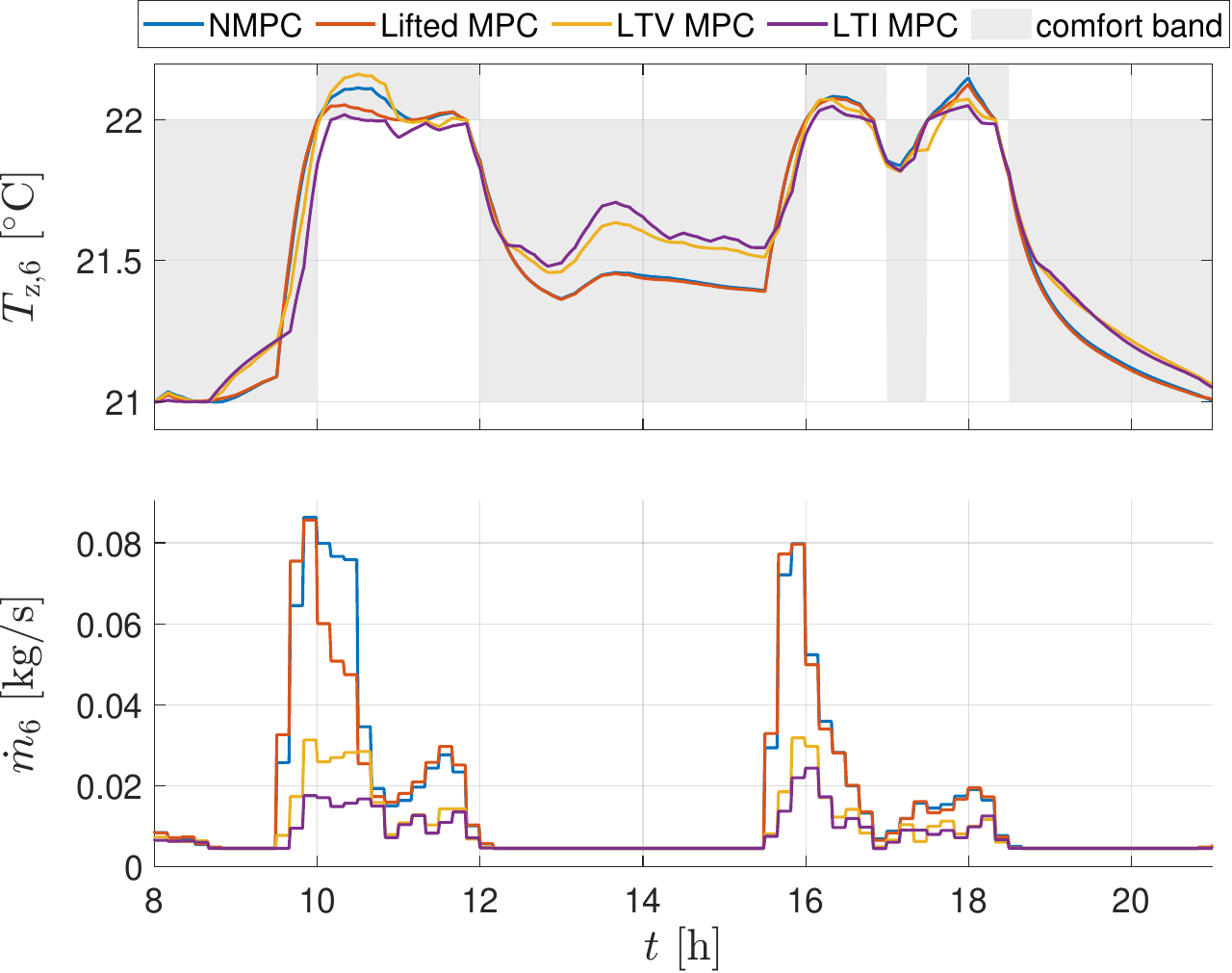}
        \caption{Zone 6, first floor.}
        \label{fig:zone6}
    \end{subfigure}
    \hfill
    \begin{subfigure}[t]{0.32\textwidth}
        \centering
        \includegraphics[width=\linewidth]{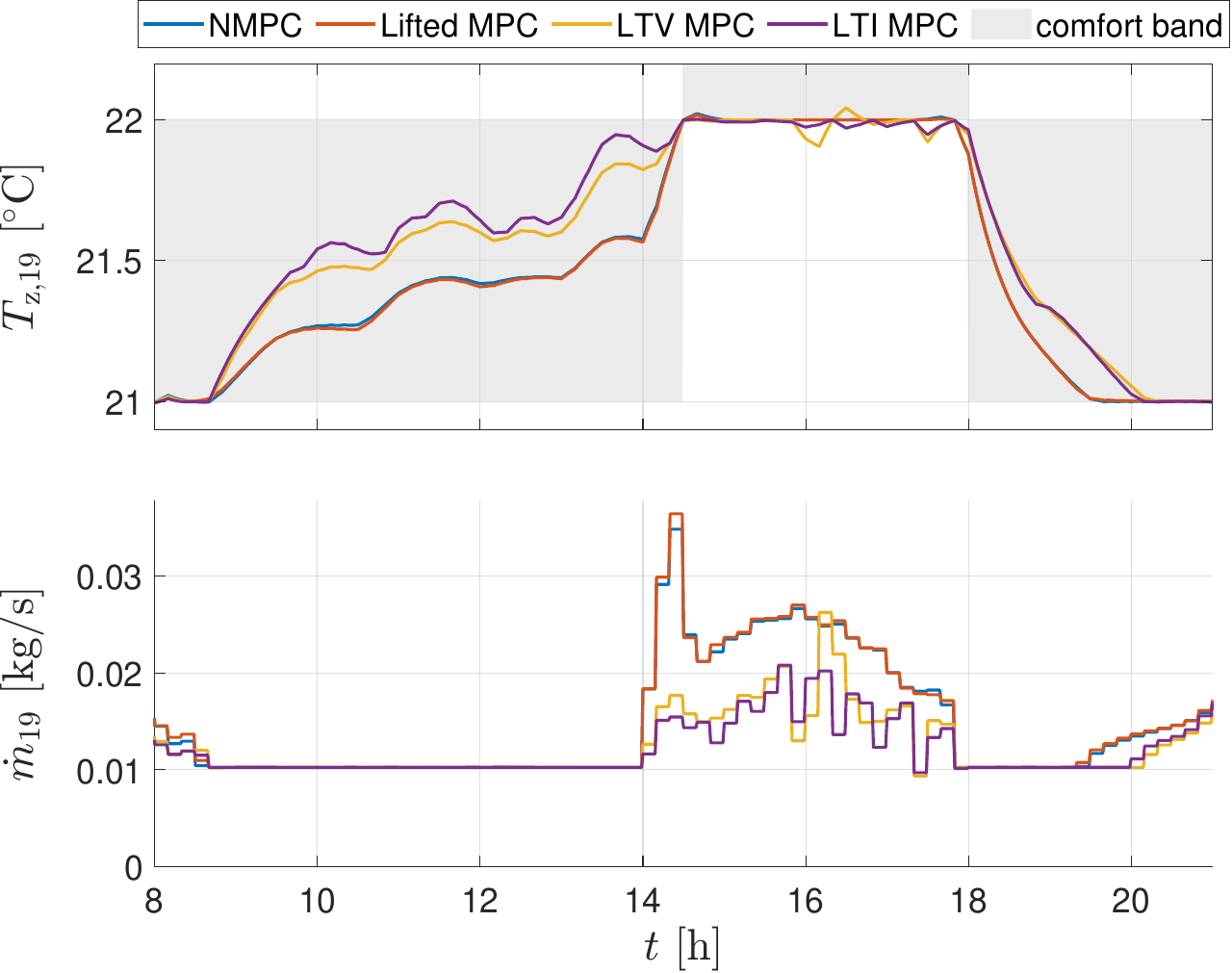}
        \caption{Zone 19, top floor.}
        \label{fig:zone19}
    \end{subfigure}

    \caption{Zone-temperature and mass-flow-rate trajectories for
    representative zones located on different floors. Lifted MPC refers to the Koopman-based MPC~\eqref{eq:MPC} with exact discretization and $\bar{P} = 4$, and LTV and LTI MPCs refer to MPC formulations based on linearizations of the original nonlinear system~\eqref{eq:ss_T}.}
    \label{fig:zone_trajectories}
\end{figure*}

\begin{figure}[!t]
    \centering
    \includegraphics[width=0.78\columnwidth]{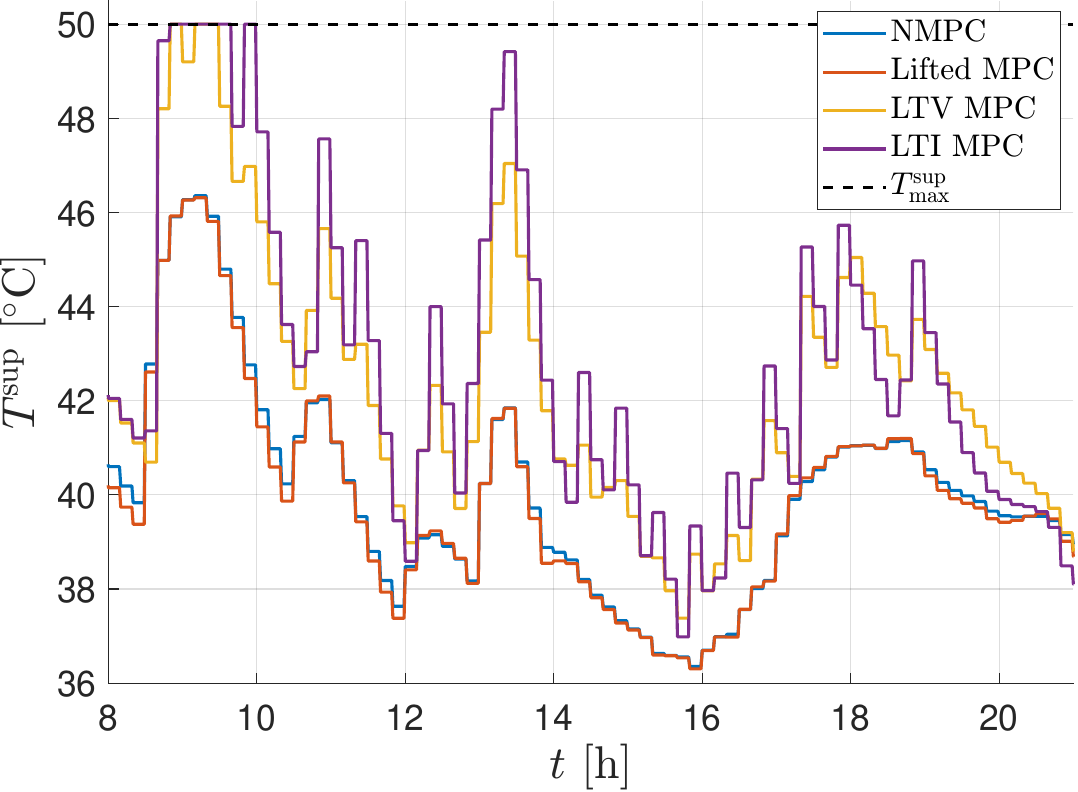}
    \caption{Supply-temperature trajectories for the considered MPC approaches.}
    \label{fig:supply_temperature}
\end{figure}

To provide a fair comparison, we consider 10 different scenarios that differ in the prediction horizon, the MPC objective weights, and the evolution of the exogenous inputs (e.g., outdoor temperature). In each scenario, the closed-loop cost is calculated from the comfort violations and energy use obtained by applying the optimized inputs to the plant, using the same weights as in the MPC objective function. The results are reported in Table~\ref{tab:mpc_comparison}, where ``vs. NMPC'' denotes the percentage difference in the closed-loop cost relative to NMPC (with walls as states), with both the average and worst-case values indicated across all scenarios. It can be observed that, for a sufficiently high truncation order, the Koopman-based MPC achieves a cost comparable to that of NMPC while being faster.\footnote{The NMPC and Koopman-based MPC are solved using IPOPT with a tolerance of \mbox{$10^{-6}$}, while the LTI and LTV approaches are solved using OSQP with \mbox{$\varepsilon_{\mathrm{abs}}=\varepsilon_{\mathrm{rel}}=10^{-4}$}.} It should be noted that the fluctuations in the cost as $\bar{P}$ increases may arise because modeling errors can sometimes favor closed-loop performance.  Moreover, both Koopman-based MPC~\eqref{eq:MPC} and the LTV approach based on the lifted model outperform the LTI and LTV approaches based on the original nonlinear system with a tuned value of $\Delta u_{\mathrm{v},\max}$, while requiring comparable computation times.

For visualization, we consider one of the scenarios with \mbox{$t_\mathrm{s}=600$~s}, \mbox{$N_\mathrm{p}=24$}, \mbox{$\rho_Q=\frac{10}{3.6\cdot 10^{6}}~\mathrm{J}^{-1}$}, and \mbox{$\lambda_i(k)=500$~K$^{-2}$} when zone \(i\) is unoccupied at time step~\(k\), \mbox{$\lambda_i(k)=10^{4}$~K$^{-2}$} when it is occupied, \mbox{$T^{\mathrm{sup}}_{\min}=25~^\circ$C}, and \mbox{$T^{\mathrm{sup}}_{\max}=50~^\circ$C}. Moreover, we use \(\dot m_{\min,i}=0.2\,\dot m_{\mathrm{nom},i}\), for all $i\in \{1, 2, ..., 20\}$, where
\[
\dot m_{\mathrm{nom},i}
=
\frac{q_{\mathrm{nom}}A^{\mathrm{floor}}_i}{\cp \Delta T_{\mathrm{nom}}},
\]
with $q_{\mathrm{nom}}=60~\mathrm{W\,m^{-2}}$ and $\Delta T_{\mathrm{nom}}=20~\mathrm{K}$. In particular,~$\dot m_{\mathrm{nom},i}$ is the design mass flow rate corresponding to the nominal heating demand of zone~$i$, with \(A^{\mathrm{floor}}_i\) denoting its floor area. Zones are considered occupied during randomly scheduled periods, which become known to the controller~30~min before they start. The temperature evolution in three zones, one located on the ground floor, one on the first floor, and one on the top floor, together with the corresponding flow rates, are depicted in Figure~\ref{fig:zone_trajectories} for the Koopman-based MPC with exact discretization and $\bar{P} = 4$, as well as the benchmark controllers. The supply temperature trajectories, each shared across all zones, are also depicted in Figure~\ref{fig:supply_temperature}.

Lastly, to demonstrate the importance of accounting for hydraulic coupling among heat emitters, we also evaluate the performance of NMPC under two simplified approaches that neglect hydraulic interactions in the system dynamics~\eqref{eq:ss_T}, using the same scenarios and parameters as above (see Table~\ref{tab:hydraulic_comparison}). The first approach ignores the hydraulic modeling in Section~\ref{sec:hydraulic_model}  and assume independent linear valve characteristics, i.e., $\dot{m}_i = u_{\mathrm{v},i} \dot{m}_{\mathrm{nom},i}$. In the second approach, we assume a fixed pressure drop $\Delta \tilde{p}$ across the heat emitters, computed from \eqref{eq:delta-p-2} by setting all valve positions equal to~$0.4$. Then, the actual flow rate is computed from $\Delta \tilde{p} = \kappa_i \dot{m}_i^2$, where $\kappa_i$ is obtained using~\eqref{eq:hydraulic_resistance}, so the valves also act independently. The valve position of $0.4$ is selected by comparing different values and evaluating the resulting performance. The results in Table~\ref{tab:hydraulic_comparison} show that accounting for hydraulic coupling improves the performance by at least~$10.9\%$.

\begin{table}[!tb]
    \centering
    \caption{Effect of considering hydraulic balancing on closed-loop performance.}
    \label{tab:hydraulic_comparison}
    \renewcommand{\arraystretch}{1.15}
    \setlength{\tabcolsep}{4pt}
    \begin{tabular}{lccc}
        \hline
        \textbf{Controller}
        & Cost
        & \textbf{vs. NMPC}
        & \textbf{\shortstack{Mean comp.\\time [s]}}
        \\
        \hline
        NMPC
        & \textbf{5490}
        & \textbf{--}
        & 7.14
        \\
        NMPC, fixed $\Delta \tilde{p}$
        & 6091
        & $+10.9\%$
        & 6.33
        \\
        NMPC, linear valves
        & 19355
        & $+252.5\%$
        & 2.81
        \\
        \hline
    \end{tabular}
\end{table}

\section{Conclusion and Future Work} \label{sec:conclusion}
In this paper, we addressed the control of hydronic heat emitters, a problem characterized by nonlinear heat-transfer dynamics and nonlinear hydraulic coupling. Using Koopman theory, we proposed an infinite-dimensional set of observables that yields an exact bilinear representation of the nonlinear model in the lifted space. For practical implementation, a finite-dimensional truncation of these observables is considered, and it is shown that the approximation error with respect to the nonlinear system is bounded and decreases with the truncation order, while the trajectories error of the truncated model remain bounded over a finite horizon, a property that is essential for control. We then incorporated the resulting bilinear model into an MPC framework to provide thermal comfort while reducing building energy use. Through a numerical case study, we showed that the proposed Koopman-based MPC achieves performance close to that of nonlinear MPC while being faster, and also outperforms LTI and LTV MPC formulations while maintaining computation times close to those of the linear approaches.

As future work, we plan to implement the proposed framework in a real building and investigate data-driven identification of the lifted bilinear model using real-world~data.

\appendix

This appendix first presents a proposition that extends
Corollary~\ref{cor:forward}. Next, we give
the proofs of Propositions~\ref{prop:trunc_bound} and~\ref{prop:trunc_tray}, and present a method this article uses for discretization.

\subsection{Equivalence of the lifted and original trajectories}\label{sec:equivalence_trajectories}

For a real sequence $v = (v_p)_{p \in \mathbb{N}}$, let
$\|v\|_{\ell^\infty} := \sup_{p \in \mathbb{N}} |v_p|$, and denote by
$\ell^\infty$ the Banach space of sequences with $\|v\|_{\ell^\infty} < \infty$. Also, define variable $\zeta_p$ as $\zeta_p = \dot m^{p-1}\chi_p$, hence $\zeta_p = u_\mathrm{T}-x_i \leq \Delta_{\max}$ for all $p \in \mathbb{N}$ (recall~\eqref{eq:Delta_max}). 

\begin{proposition}
\label{prop:equivalence}
Let $\psi:\Xi\to z$ be the global lifting, and $\mathcal M:=\psi(\Xi)$ be the
embedding manifold, where $\Xi$ is the set of $\xi=(x,\mu,u_T)$ with
$\mu_i\in[\log\dot m_{\min,i},\log\dot m_{\max,i}]$ for all $i\in \{1, 2, ..., n_\mathrm{z}\}$. Let $z(0)\in\mathcal M$,
and $\tilde u(t)$ be an input such that $\xi(t)\in\Xi$ for all $t$. 
Then, 
\begin{enumerate}
  \item[(a)] $\psi$ is injective, with inverse
  \begin{equation}
    x_i=\theta_{i,1},\qquad \dot m_i=\phi_{i,0},\qquad u_T=\theta_{i,1}+\chi_{i,1},
    \label{eq:inverse}
  \end{equation}
  for all $i\in \{1, 2, ..., n_\mathrm{z}\}$. 
  \item[(b)] If $\xi(t)$ solves the original nonlinear system \eqref{eq:nl_dynamics}, then $z(t) = \psi(\xi(t))$ solves the lifted system \eqref{eq:final_z_tilde}.
  \item[(c)] If $z(t)$ solves \eqref{eq:final_z_tilde} with $z(0)\in\mathcal M$
  and $\zeta(t)\in\ell^\infty$, then $z(t)=\psi(\xi(t))$ and $\xi(t)=\psi^{-1}(z(t))$.
\end{enumerate}
\end{proposition}

\begin{proof}
(a) follows directly from the definition of $\psi$, and (b)
is the forward direction established in Corollary~\ref{cor:forward}. The remaining task is (c), which
requires that the infinite lifted system has \emph{at most one} solution. This is
not straightforward for two reasons: \eqref{eq:final_z_tilde} contains terms
$(1-p)\tilde u_{\mu,i}$, which are unbounded in $p$, and also
infinite sums (see \eqref{eq:matrix_omega}). To prove~(c), we consider the following steps. With a slight abuse of notation, index $i$ is dropped for clarity.

\medskip
\noindent\textit{Step 1: Solve decoupled scalar equations for some of the observables.}
Variables $\phi_p$ and $s_p:=\theta_p+\chi_p=\phi_p u_T$ each satisfy a
scalar linear ODE that does not involve the other variables. Specifically, 
\begin{equation}
  \dot\phi_p=(1-p)\,\tilde u_\mu\,\phi_p,
  \qquad
  \dot s_p=(1-p)\,\tilde u_\mu\,s_p+\phi_p\,\tilde u_T .
\end{equation}
Both have unique solutions and can be solved explicitly for a given $\tilde u$. In particular,
\begin{equation}
  \phi_0(t)=\dot m(t) =\phi(0)\exp\!\Big(\int_0^t\tilde u_\mu\,\mathrm d\tau\Big).
\end{equation}

\noindent\textit{Step 2: Remove the unbounded terms by rescaling.}
Substituting $\theta_p=s_p-\chi_p$ into \eqref{eq:dchi} gives
\begin{equation}\label{eq:chi_s}
 \begin{split}
        \dot\chi_{p} 
         &=\left((1\!-\!p)\tilde u_\mu +a\right)\chi_p
  +b\sum_{n=1}^\infty\alpha_n\,\chi_{p+n-1}+h_p(t),
 \end{split}
 \end{equation}
where $h_p:=\phi_p\tilde u_T-a\,s_p-\phi_p\,w$. 
Moreover, multiplying~\eqref{eq:chi_s} by $\dot m^{p-1}$ results in
\begin{equation*}
 \begin{split}
       &\dot m^{p-1} \dot\chi_{p}  \\
       &= \dot \zeta_p - (p-1)\tilde u_\mu \dot m^{p-1} \chi_p = \dot \zeta_p - (p-1) \zeta_p \tilde u_\mu   \\
  &=(1-p) \zeta_p \tilde u_\mu +a \zeta_p
  +b\sum_{n=1}^\infty\alpha_n \dot m^{1-n} \zeta_{p+n-1}+ \dot m^{p-1} h_p, 
 \end{split}
 \end{equation*}
 where we have used the definition of $\zeta_p$ and $\frac{\mathrm d}{\mathrm dt}\dot m^{1-p}=\frac{\mathrm d}{\mathrm dt}\phi_p = (1-p)\tilde u_\mu\dot m^{1-p}$.
 The term $(1-p) \zeta_p \tilde u_\mu$ in the first and last equality cancel exactly, leaving
\begin{equation}
\begin{split}
\dot\zeta_p&=a\,\zeta_p+b\sum_{n=1}^\infty\alpha_n\,\dot m^{1-n}\,\zeta_{p+n-1}+\hat h_p \\
  &=\;a\,\zeta_p+b\,\big(B(t)\zeta\big)_p+\hat h_p,
  \label{eq:zeta}
\end{split}
\end{equation}
with forcing $\hat h_p:=\tilde u_T-a\,u_T-w$. The forcing no longer depends on $p$ and is bounded. Writing $(S\zeta)_p:=\zeta_{p+1}$
for the forward shift (so $\|S\|_{\ell^\infty}=1$),   the operator $B(t)$ and its
norm are
\begin{equation}
\begin{split}
     &B(t)=\sum_{n=1}^\infty \alpha_{n}\,\dot m(t)^{1-n}S^{n-1},
  \\
  &\|B(t)\|_{\ell^\infty}\le\sum_{n=1}^\infty \frac{c^n}{n!}\,\dot m(t)^{1-n} 
  =\dot m(t)\big(e^{c/\dot m(t)}-1\big), 
  \end{split}
\end{equation}
where argument $t$ has been explicitly included to emphasize the time dependency. Since $\dot m(t)\in[\dot m_{\min},\dot m_{\max}]$ for all $t$, $\|B(t)\|_{\ell^\infty}\le \dot m_{\max}\,e^{c/\dot m_{\min}} = \bar B$. Hence, $B(t)$ is a bounded linear operator, and the bound $\bar B$ does not depend on $t$. 

\vspace{5pt}
\noindent\textit{Step 3: Conclude by uniqueness for all observables.}
Equation \eqref{eq:zeta} is a linear, time-varying ODE on the Banach space
$\ell^\infty$, with a bounded operator and bounded forcing. By the
Picard--Lindel\"of theorem in Banach spaces \cite[Ch.~5, Thm.~5.1]{pazy2012semigroups}, it has a
unique solution $\zeta(t) = (\zeta_0(t), \zeta_1(t), ...)$ for every initial condition $\zeta(0)\in\ell^\infty$. Considering Step 1, this also implies that there are unique solutions $\chi_p(t)$ and $\theta_p(t)$ for all $p \in \mathbb{N}$. 

It remains to check that the lift of the original trajectory is such a solution.
Again, let $\xi(t)$ solve the original system and consider its lift $\psi(\xi(t))$,
which solves the lifted system by (a). On $\mathcal M$,  we have
$\chi_p=\dot m^{1-p}\zeta_p$, where $\zeta_p=u_\mathrm{T}-x$, and $\|\zeta(t)\|_{\ell^\infty}\le\Delta_{\max}$. Thus
$\psi(\xi(t))$ is an $\ell^\infty$ solution, and by uniqueness it is the only one
starting from $\psi(\xi(0))$. Any solution $z(t)$ with
$z(0)=\psi(\xi(0))$ and $\zeta(t)\in\ell^\infty$ therefore equals
$\psi(\xi(t))$, and $\xi(t)$ is recovered through the inverse~\eqref{eq:inverse}. This proves (c).
\end{proof}

\subsection{Proofs of Propositions~\ref{prop:trunc_bound} and~\ref{prop:trunc_tray}}\label{app:proofs}

With a slight abuse of notation, we again omit index $i$ for clarity in what follows. 

\begin{proofof}{Proposition~\ref{prop:trunc_bound}}
Using $\chi_{p+n-1} = \dot m^{\,2-p-n}(u_\mathrm{T}-x)$ and introducing the definition of $\alpha_{n}$, we have that
\begin{equation}
\begin{split}
    r_{p}^{(N(p))} &= (u_\mathrm{T}-x)  \sum_{n=N(p)+1}^{\infty} \frac{(-c)^n}{n!}\,\dot m^{\,2-p-n}\\
    &= (u_\mathrm{T}-x)\,\dot m^{\,2-p} \sum_{n=N(p)+1}^{\infty} \frac{1}{n!}\left(\frac{-c}{\dot m}\right)^{\!n}.
\end{split}
\end{equation}
The sum in the last equality is the remainder of the Taylor series of $e^{y}$ for $y = -c/\dot m$, truncated after $N(p)$ terms, and satisfies the standard bound
\begin{equation}
    \left|\sum_{n=N+1}^{\infty} \frac{y^n}{n!}\right| \le \frac{|y|^{N+1}}{(N+1)!}\,e^{|y|}.
\end{equation}
Since $\dot m \ge \dot m_{\min} > 0$, we have $|y| \le \mathrm{NTU}_{\max}$, and the right-hand side above is nondecreasing in $|y|$. Moreover, $|u_\mathrm{T} - x| \le \Delta_{\max}$ and $\dot m^{\,2-p} \le \bar M_p$. Combining these bounds gives the inequality in~\eqref{eq:trunc_bound}.

Finally, note that $p \ge 2$ implies $2-p \le 0$, so $\bar M_p = \dot m_{\min}^{\,2-p}$. Since $N(p) = \bar P - p + 1$,
\begin{equation}
   \bar M_p\,\mathrm{NTU}_{\max}^{\,N(p)+1}
   = \dot m_{\min}^{\,2-p}\,\frac{c^{\,\bar P-p+2}}{\dot m_{\min}^{\,\bar P-p+2}}
   = \dot m_{\min}^{-\bar P}\, c^{\,\bar P-p+2}.
\end{equation}
Substituting this gives the equality in~\eqref{eq:trunc_bound}.
\end{proofof}

\begin{proofof}{Proposition~\ref{prop:trunc_tray}}
Consider Proposition~\ref{prop:equivalence}, and in particular~\eqref{eq:zeta}. For a certain truncation order $\bar P$,~\eqref{eq:zeta} reads
\[
\dot{\zeta}_p^{(\bar P)} = a\zeta^{(\bar P)}_p + b (B^{(\bar P)}(t)\zeta^{(\bar P)})_p + \hat{h}_p,
\]
where $B^{(\bar P)}$ retains the shifts with $p+n-1\le \bar P$. The error
$e_p \coloneqq \zeta_p-\zeta_p^{(P)}$ on the retained indices satisfies $e(0)=0$ and
\begin{equation}
   \begin{split}
      & \dot{e}_p = \bigl(a+b B^{(\bar P)}(t)\bigr)e + b \tau_p(t), \\
\qquad
&\tau_p(t) \coloneqq \bigl( \bigl(B(t)-B^{(\bar P)}(t)\bigr)\zeta(t)\bigr)_p,
   \end{split} 
\end{equation}
where $|\tau_p(t)| = \dot m(t)^{p-1}\bigl| r_{p}^{(N(p))}(t)\bigr|$
by~\eqref{eq:remainder}, so the forcing term is the remainder rescale.  With
$L := |a| + b\,\dot{m}_{\max} e^{\mathrm{NTU}_{\max}} \geq \|a + bB^{(\bar{P})}(t)\|$
and $\bar{\tau} := \sup_{t,\, p \leq \bar{P}} \dot{m}(t)^{p-1} \bigl|r_{p}^{(N(p))}(t)\bigr|$,
finite by~Proposition~\ref{prop:trunc_bound},
\[
  \frac{d}{dt}\|e\|_{\ell^\infty} \leq L\,\|e\|_{\ell^\infty} + b\,\bar{\tau},
  \quad \|e(0)\|_{\ell^\infty} = 0.
\]
By Gr\"onwall's inequality~\cite[Lemma A.1]{khalil2002nonlinear}, we obtain
\[
 \|e(t)\|_{\ell^\infty} \leq \frac{b\,\bar{\tau}}{L}\bigl(e^{Lt} - 1\bigr).\]
Finally, since $x - x^{(\bar{P})} = -e_1$,
this bounds the zone temperature prediction error over the horizon. 
\end{proofof}

\subsection{Discretization of the Bilinear System~\eqref{eq:disc_ztilde}}\label{app:disc}

Let us simplify the notation of the continuous-time bilinear model in~\eqref{eq:final_z_tilde} by expressing it for each zone, since the zones are decoupled, as
\begin{equation}
\dot z_i
=
\left(
A_i
+\dot u_\mathrm{T}N_\mathrm{T}
+\dot\mu_i N_\mu
+w_i N_w
\right)z_i,
\label{eq:zone_ct}
\end{equation}
where $z_i\in\mathbb{R}^{3(\bar{P}+1)}$, $u_\mathrm{T},\mu_i,w_i\in\mathbb{R}$, and $A_i,N_\mathrm{T},N_\mu,N_w\in\mathbb{R}^{3(\bar{P}+1)\times 3(\bar{P}+1)}$. For the discretization, the inputs are applied with a zero-order hold. In particular, the supply temperature and the logarithmic flow rates change only at the sampling instants $t=k t_\mathrm{s}$, by $\Delta u_\mathrm{T}(k)=t_\mathrm{s}\tilde u_\mathrm{T}(k)$ and $\Delta\mu_i(k)=t_\mathrm{s}\tilde u_{\mu,i}(k)$, and are held constant in between. Moreover, the exogenous input $w_i$ is held at its forecast mean $w_i(k)$. As a result, the terms with $N_\mathrm{T}$ and $N_\mu$ in~\eqref{eq:zone_ct} act only at $t=k t_\mathrm{s}$, and integrating across this input change gives
\begin{align}
z_i^{+}(k)&=e^{\Delta u_\mathrm{T}(k)N_\mathrm{T}+\Delta\mu_i(k)N_\mu}z_i(k) \nonumber \\
&=\big(I+\Delta u_\mathrm{T}(k)N_\mathrm{T}\big)S\big(\Delta\mu_i(k)\big)z_i(k),
\end{align}
where $z_i(k)$ and $z_i^{+}(k)$ denote the states just before and after the change, and $S(\Delta\mu_i)=\mathrm{diag}\big(e^{(1-p)\Delta\mu_i}\big)$, since $N_\mu$ is diagonal, $N_\mathrm{T}^2=0$ and $N_\mu N_\mathrm{T}=N_\mathrm{T}N_\mu$. Over the interval, all inputs are constant, and since $N_wA_i=0$ and $N_w^2=0$, the solution of $\dot z_i=\big(A_i+w_i(k)N_w\big)z_i$ yields
\begin{equation}
z_i(k+1)=\big(A_{\mathrm{d},i}+\Gamma_i N_w w_i(k)\big) z_i^{+}(k),
\label{eq:zone_dt}
\end{equation}
with $A_{\mathrm{d},i}=e^{A_i t_\mathrm{s}}$ and $\Gamma_i=\int_0^{t_\mathrm{s}}e^{A_i\tau}\mathrm{d}\tau$. Since the two parts act at different times, the discretized model \eqref{eq:zone_dt} is exact.

\section*{Acknowledgment}
The authors would like to thank Shalika Walker and Christian Portilla for their support in facilitating this research.

\bibliographystyle{IEEEtran}
\bibliography{IEEEabrv}
\end{document}